\documentclass[eqsecnum,aps,prd,nofootinbib]{revtex4}
\usepackage[dvips]{graphicx}

\usepackage{amsmath}
\usepackage{enumerate}
\usepackage{mathrsfs}
\usepackage{amsfonts}
\usepackage{cancel}

\newcommand{\I}{{\mathcal I}}

\newcommand{\F}{{\mathcal F}}
\newcommand{\tg}{{\tilde g}}

\newcommand{\tD}{{\tilde D}}
\newcommand{\tn}{{\tilde n}}
\newcommand{\tgam}{{\tilde \gamma}}
\newcommand{\thh}{{\tilde h}}
\newcommand{\tP}{{\tilde P}}

\renewcommand{\d}{{\partial }}

\def\varQ#1#2{\frac{\delta #1}{\delta #2}}

\def\eps{\epsilon}

\def\d{\partial}

\def\cD{{\cal D}}

\def\cI{{\cal I}}

\def\cL{{\cal L}}
\def\cM{{\cal M}}
\def\cN{{\cal N}}

\def \eps{{\epsilon}}

\begin{document}
\preprint{}
\title{Supergravity at the boundary of AdS supergravity}

\author{Aaron J. Amsel and Geoffrey Comp\`ere}
\affiliation{%
Department of Physics \\ University of California, Santa Barbara \\ Santa Barbara, CA 93106, USA
}

\begin{abstract}
We give a general analysis of AdS boundary conditions for spin-3/2 Rarita-Schwinger fields and investigate boundary conditions preserving supersymmetry for a graviton multiplet in AdS$_4$.
Linear Rarita-Schwinger fields in AdS$_d$ are shown to admit mixed Dirichlet-Neumann boundary conditions when their mass is in the range $0 \leq |m| < 1/2l_{AdS}$. We also demonstrate that mixed boundary conditions are allowed for larger masses when the inner product is ``renormalized'' accordingly with the action. We then use the results obtained for $|m|  = 1/l_{AdS}$ to explore supersymmetric boundary conditions for $\cN = 1$ AdS$_4$ supergravity in which the metric and Rarita-Schwinger fields are fluctuating at the boundary. We classify boundary conditions that preserve boundary supersymmetry or superconformal symmetry. Under the AdS/CFT dictionary, Neumann boundary conditions in $d=4$ supergravity correspond to gauging the superconformal group of the 3-dimensional CFT describing M2-branes, while $\cN = 1$ supersymmetric mixed boundary conditions couple the CFT to $\cN = 1$ superconformal topologically massive gravity.

\end{abstract}


\maketitle

\tableofcontents

\section{Introduction}

Supergravity theories can be understood as low energy approximations to string theory. This has led via the AdS/CFT correspondence (see the review \cite{Aharony:1999ti}) to the recognition that supergravity theories are dual to conformal field theories in the planar limit and at strong coupling. In the usual formulation of the AdS/CFT correspondence, the leading order coefficients in the radial expansion of bulk fields near the boundary are held fixed.  These coefficients, hereafter referred to as ``boundary fields,'' are then interpreted as sources for dual operators in the boundary CFT.

Deforming the boundary conditions corresponds on the field theory side to multi-trace deformations of the CFT action \cite{Witten:2001ua,Berkooz:2002ug}. A systematic analysis of boundary conditions for linear matter fields around AdS has been performed for spin-0 fields \cite{Breitenlohner:1982jf,Breitenlohner:1982bm,Klebanov:1999tb,Ishibashi:2004wx}, spin-1/2 fields \cite{Breitenlohner:1982jf,Breitenlohner:1982bm,Amsel:2008iz}, and spin-1 fields \cite{Breitenlohner:1982jf,Breitenlohner:1982bm,Ishibashi:2004wx,Marolf:2006nd}.  For bulk scalar supermultiplets in AdS, these results were used to classify the supersymmetric multi-trace deformations of the corresponding dual CFTs in \cite{Hollands:2006zu,Amsel:2008iz}. In this work, we will be interested in supersymmetric boundary conditions for the graviton multiplet, which contains a spin-3/2 Rarita-Schwinger fermion. Rarita-Schwinger fields in AdS with the standard Dirichlet boundary conditions have been extensively studied \cite{Deser:1984py,Volovich:1998tj,Koshelev:1998tu,Corley:1998qg,Rashkov:1999ji,Matlock:1999fy,Rychkov:1999zp}, though no complete treatment including more general boundary conditions has been developed. Hence, the first aim of this paper is to fill this gap.

Using the standard definition of the symplectic structure, it is generally accepted that the leading mode of the graviton close to the conformal AdS boundary is ``non-normalizeable.'' However, a detailed examination of linearized gravitons around AdS \cite{Ishibashi:2004wx} indicated that a more interesting treatment is possible for some graviton modes in $d = 4,5,6$ spacetime dimensions. This hint motivated the work \cite{Compere:2008us} where, perhaps surprisingly, it was shown that alternative ``Neumann-type'' boundary conditions (in which the boundary metric is dynamical) exist for the full non-linear gravity theory in any dimension (see also related comments in \cite{Leigh:2003ez,Marolf:2006nd,Leigh:2007wf,deHaro:2008gp}). Varying the ``non-normalizeable'' part of the bulk metric turns out to be allowed once the symplectic structure is renormalized accordingly with the action. In short, all infinities appearing in the standard symplectic structure are canceled by the contributions of the counterterms.

The second aim of this paper is to analyze boundary conditions for Rarita-Schwinger fields using the renormalized symplectic structure.  Once the renormalization has been performed, modes that were previously considered as ``non-normalizeable'' are now allowed to fluctuate. This step will allow us later in the paper to extend the Neumann boundary conditions for gravity to supergravity.  For minimal supergravity in AdS$_4$, the standard Dirichlet boundary conditions were shown to preserve supersymmetry on the boundary in \cite{Hollands:2006zu}.

Allowing the boundary metric to fluctuate corresponds to varying the metric on which the CFT lives, i.e., to considering the induced gravity of the dual CFT. The resulting boundary gravity theory is non-local in the metric or, equivalently, admits an infinite expansion in higher curvature terms. In the simple example of two dimensions, conformal invariance dictates that the effective action of any CFT is the Polyakov action \cite{Polyakov:1987zb}.
Higher-derivative gravity theories are generically plagued with a lack of unitarity \cite{Fradkin:1985am}. However, quite remarkably, the induced gravity theories obtained in odd boundary dimensions $(d-1)$ were shown to be ghost and tachyon free around a flat boundary\footnote{For even dimensional boundaries, the theories contain both tachyons and ghosts around flat space. However, the theory may be still be stable in some parameter region around de Sitter space, which may have consequences for certain inflation models \cite{Hawking:2000kj,Hawking:2000bb}.} \cite{Compere:2008us}. The simplest such theory is the dual to four-dimensional Einstein gravity with negative cosmological constant.  As the precise mapping between strongly coupled CFTs and gravitational theories generally requires a large amount of bulk supersymmetry, it is a natural step to extend the pure gravity results of \cite{Compere:2008us} to $\cN = 1$ AdS$_4$ supergravity \cite{Deser:1976eh,Townsend:1977qa}\footnote{Boundary conditions imposed at a \emph{finite} value of the radial coordinate have been studied recently in supergravity with zero cosmological constant \cite{vanNieuwenhuizen:2005kg,vanNieuwenhuizen:2006pz,Belyaev:2008ex}. In this current work, the boundary is at infinity.}. To analyze boundary conditions for this theory, we follow the holographic renormalization procedure \cite{deHaro:2000xn,Bianchi:2001kw,Bianchi:2001de}.

The bulk fields of supergravity theories with negative cosmological constant generically induce conformal supergravity multiplets at the conformal boundary. In particular, it was noticed soon after the emergence of AdS/CFT that the gauged $d=5$, $\cN = 8$ supermultiplet in the bulk induces the $d=4$, $\cN = 4$ conformal supermultiplet at the boundary \cite{Ferrara:1998ej,Liu:1998bu}. In $d = 3$, $(p,q)$-multiplets generate two-dimensional $(p,q)$-conformal supermultiplets at the boundary  (for $p,q \leq 2$) \cite{Nishimura:1998ud}, while in the less studied gauged $d=6$ $F(4)$ supergravity, the boundary fields form a $\cN = 2$ superconformal multiplet \cite{D'Auria:2000ad,Nishimura:2000wj}. One can also argue that gauged $d=4$, $\cN = 8$  supergravity will induce the $d=3$, $\cN = 8$ conformal supermultiplet at the boundary. Indeed, it has been shown recently that $\cN = 8$ superconformal gravity can be coupled to Bagger-Lambert-Gustavsson theory \cite{Gran:2008qx}, which is closely related \cite{Antonyan:2008jf} to the CFT dual of M-theory on $AdS_4 \times \mathbb C^4/\mathbb Z_k$ \cite{Aharony:2008ug}. In this paper, we will show that at least the $\cN = 1$ conformal supermultiplet \cite{vanNieuwenhuizen:1985cx} is induced at the conformal boundary of minimal supergravity.

The outline of this paper is as follows. Section \ref{RS} is devoted to linear Rarita-Schwinger fields in $d$ spacetime dimensions. After reviewing general properties of spin-3/2 fields in AdS, we analyze admissible boundary conditions with respect to both the standard symplectic structure and the renormalized symplectic structure. In the last subsection, we review the standard Euclidean propagator for spin-3/2 fields with Dirichlet boundary conditions.  We then determine the corresponding propagator in the Neumann theory and discuss the presence of tachyons. In section \ref{nonlinear}, we implement the holographic renormalization procedure for $d=4$ minimal supergravity. In particular, this requires solving the equations of motion asymptotically and finding the allowed asymptotic gauge transformations. We then discuss renormalizing the action by adding certain boundary counterterms and show that the action is invariant under supersymmetry, special conformal supersymmetry, and Weyl transformations at the boundary. Section \ref{disc} contains a summary of our results and a discussion of unitarity in theories with Neumann boundary conditions.

\subsection{Preliminaries}
\label{prelim}

Before restricting to any specific theory, we review the general definition of the symplectic structure.  This formalism will be applied to the theory of spin-3/2 fields  linearized about AdS in section \ref{RS} and to AdS supergravity in section \ref{nonlinear}.   Conventions used throughout this work are summarized in appendix  \ref{sec:conventions}.

In AdS spacetimes, one must be careful to choose boundary conditions that lead to a well-defined bulk theory.  This means in particular that the bulk symplectic structure is both finite and conserved.  In general, the first variation of a Lagrangian density (written as a $d$-form) for a set of fields $\phi$ can be expressed as
$
\label{dL} \delta {\bf L} = {\bf E} \cdot \delta \phi +
d\boldsymbol{\theta} \,, $ where the equations of motion of the theory are $\mathbf{E} = 0$,
and $d\boldsymbol{\theta}$ corresponds to surface terms that would
arise from integrating by parts. Now, let $\Sigma$ be a constant time hypersurface with unit normal $t^\mu$.  Then, for linearized solutions $\delta_{1,2} \phi$, the standard symplectic structure on $\Sigma$ is
\begin{equation}
\label{canonstructure}
\sigma_{\Sigma}(\phi; \delta_1 \phi, \delta_2 \phi) = \int_\Sigma
\boldsymbol{\omega}(\phi; \delta_1 \phi, \delta_2 \phi)\,,
\end{equation}
where we have defined the standard symplectic current (a $(d-1)$-form ) as
\begin{equation}
\label{current} \boldsymbol{\omega}(\phi; \delta_1 \phi, \delta_2
\phi) = \delta_1 \boldsymbol{\theta}(\phi; \delta_2 \phi) - \delta_2
\boldsymbol{\theta}(\phi; \delta_1 \phi).
\end{equation}
For scalar fields, $\sigma_\Sigma(\phi; \delta_1 \phi, \delta_2 \phi)$ is just the familiar Klein-Gordon inner product.

It is well-known that the symplectic potential, $\boldsymbol{\theta}$, admits an ambiguity in its definition given by $\boldsymbol{\theta} \to \boldsymbol{\theta}+ d \mathbf{B}$, where $\mathbf{B}$ is an arbitrary $(d-2)$-form. This leads directly to a corresponding ambiguity in the definition of the symplectic current \eqref{current}.  However, it has been suggested in \cite{Compere:2008us} that this ambiguity can be fixed once the action has been ``renormalized.'' This refers to the fact that the subtraction of divergences in the action by diffeomorphism-invariant boundary terms $\int_{\I} {\bf L}_{ct}[\phi_*]$ is required in order to render the action finite on-shell and to define the boundary stress-tensor \cite{Balasubramanian:1999re,deHaro:2000xn,Papadimitriou:2004ap}. Here we denote the AdS conformal boundary by $\I$ and the induced fields on the boundary collectively by $\phi_*$.  Now, these boundary terms generally contain time derivatives of the boundary fields and therefore should contribute to the symplectic structure\footnote{In the case of Einstein gravity, the Gibbons-Hawking and cosmological boundary terms do not provide a contribution to the symplectic structure, while the boundary Einstein and higher curvature actions do contribute \cite{Compere:2008us}.}. In addition, non-minimal (finite) terms may be added to the boundary action and should also contribute to the boundary dynamics. To see this, we first define $\boldsymbol{\theta}_{ct}[ \phi_*]$ by the variation formula $\delta {\bf L}_{ct} = \varQ{{\bf L}_{ct}}{\phi_*^i}\delta \phi_*^i + d_*\boldsymbol{\theta}_{ct}[\phi_*;\delta \phi_*]$, where $d_*$ is the induced exterior derivative on surfaces of constant radial coordinate. We then define the corresponding renormalized symplectic current as
\begin{equation}
\boldsymbol{\omega}_{ren} = \boldsymbol{\omega} + d \boldsymbol{\omega}_{ct} \label{renomega} \, ,
\end{equation}
where $\boldsymbol{\omega}_{ct} = \delta_1 \boldsymbol{\theta}_{ct}[\phi_* ; \delta_2 \phi_*] - (1 \leftrightarrow 2)$, and $d\boldsymbol{\omega}_{ct}$ is an arbitrary smooth extension of $d_* \boldsymbol{\omega}_{ct}$ away from the boundary. For odd dimensional boundaries, this definition will be independent of the choice of radial foliation. However, for even dimensional boundaries the renormalized symplectic structure may depend explicitly on the radial foliation because of the conformal anomaly.

In general, varying the renormalized action yields a finite expression of the form $\delta S = \int_\I d^{d-1}x \,\pi_*^i \delta \phi^*_i$, which defines the fields $\pi_*^i$ that are conjugate to $\phi^*_i$. The renormalized symplectic flux through the boundary is then given by the finite expression
\begin{equation}
\F_{ren} = \int_\I \boldsymbol{\omega}_{ren} = \int_\cI d^{d-1}x\,\left( \delta_1 \pi^i_*\; \delta_2 \phi^*_i - (1 \leftrightarrow 2)\right)  \,. \label{flux}
\end{equation}
Note that since $\boldsymbol{\omega}$ and $\boldsymbol{\omega}_{ren}$ differ only by a boundary term, it is possible to deduce the integrand in \eqref{flux} from the original Lagrangian $L$. However, the counterterms are needed to remove the divergences occurring at the boundaries $\partial \cI$. In order to ensure that the symplectic structure is conserved, i.e. that $\F_{ren} = 0$, we have to impose either Dirichlet ($\phi^*_i$ fixed), Neumann ($\pi^i_*$ fixed), or mixed ($\pi^i_* = \varQ{W[\phi^*_i]}{\phi^*_i}$ for some arbitrary function $W$) boundary conditions.

Let us consider a spacetime region bounded by two constant time slices $\Sigma_1, \Sigma_2$ whose boundaries $\partial \Sigma_1 \equiv S_1$, $\partial \Sigma_2 \equiv S_2$ are $S^{d-2}$ spheres\footnote{Our conventions for the orientation of surfaces in the application of Stokes's Theorem match those given in appendix B of \cite{Waldbook}.}. Since the symplectic current is closed on-shell, we can write the flux in the form
\begin{equation}
\F_{ren} = \sigma_{\Sigma_2, ren} - \sigma_{\Sigma_1, ren} \,,
\end{equation}
where the symplectic structure associated with a $ t= constant$ hypersurface $\Sigma$ is given by
\begin{equation}
\label{renormstructure}
\sigma_{\Sigma, ren}(\phi; \delta_1 \phi, \delta_2 \phi) = \int_\Sigma \boldsymbol{\omega}(\phi; \delta_1 \phi, \delta_2 \phi) - \int_{\partial \Sigma} \boldsymbol{\omega}_{ct}(\phi_*; \delta_1 \phi_*, \delta_2 \phi_*) \,.
\end{equation}
We see that the renormalized symplectic structure \eqref{renormstructure} differs from the standard symplectic structure \eqref{canonstructure} only when boundary fields are allowed to vary. Now, it was found in \cite{Compere:2008us} that when $\phi$ is the metric field, this boundary contribution cancels the divergences appearing in the standard symplectic structure. This result is expected to hold in general. Below, we indeed verify explicitly the cancelation of divergences for a large range of parameters in the linear spin-3/2 theory. In that sense, the renormalized symplectic structure extends the standard definition by transforming ``non-normalizeable'' modes into ``renormalized'' modes.

\section{Rarita-Schwinger fields in AdS$_d$ background}
\label{RS}

We now consider the linearized theory of spin-3/2 fermions in AdS$_d$, i.e., the fields propagate on a fixed AdS background.   After providing general asymptotic solutions to the Rarita-Schwinger equation, we then analyze normalizeability of the solutions with respect to the standard symplectic structure and the renormalized symplectic structure.  To conclude, we briefly discuss several interesting features of the spin-3/2 boundary propagator.

\subsection{``Massive'' Rarita-Schwinger fields}
\label{massive}

We now review the massive Rarita-Schwinger equation in AdS spacetime.  The most general Rarita-Schwinger action in $d$ spacetime dimensions is
\begin{equation}
\label{rsaction}
S = N \int d^dx \sqrt{-g} \left[ \, \overline{\psi_\mu} \Gamma^{\mu \nu \lambda} D_\nu \psi_\lambda
+ m_1 \overline{\psi_\mu} \Gamma^{\mu \lambda} \psi_\lambda + m_2 \overline{\psi_\mu} \psi^\mu \right] \,,
\end{equation}
where $N$ is some constant normalization that for now we leave arbitrary.
In the above action, we take the (torsion-free) covariant derivative acting on a vector-spinor to be
\begin{equation}
D_\rho \psi_\sigma = \partial_\rho \psi_\sigma - C^\lambda_{\rho \sigma} \psi_\lambda +\frac{1}{4} \omega_\rho{}^{\hat \mu \hat \nu}(e) \Gamma_{\hat \mu \hat \nu} \psi_\sigma  \,, \label{dpsi}
\end{equation}
where $\omega_\rho{}^{\hat \mu \hat \nu}(e)$ are the rotation coefficients defined by the ``vielbein postulate''
\begin{equation}
D_\rho e^{\hat \mu}{}_{\mu} = \partial_\rho e^{\hat \mu}{}_{\mu} -C^{\lambda}_{\rho \mu} e^{\hat \mu}{}_{\lambda} + \omega_\rho{}^{\hat \mu}{}_{\hat \nu}(e) e^{\hat \nu}{}_{ \mu} = 0 \label{textbookomega}
\end{equation}
and $C^\lambda_{\rho \sigma}$ is the standard Christoffel symbol.  Note, however, that torsion terms will be required in the nonlinear supergravity theory discussed in section \ref{nonlinear}.

By varying the action \eqref{rsaction} with respect to the Rarita-Schwinger field, we obtain the equations of motion
\begin{equation}
\label{rs1}
\Gamma^{\mu \nu \lambda} D_\nu \psi_\lambda
+ m_1 \Gamma^{\mu \lambda} \psi_\lambda +m_2 \psi^\mu =0 \,.
\end{equation}
If we apply $D_\mu$ to both sides of the Rarita-Schwinger equation \eqref{rs1} and use the gamma matrix identity
$\Gamma^{\mu \nu \lambda} = \gamma^{\mu} \gamma^{\nu} \gamma^{\lambda} - g^{\mu \nu} \gamma^\lambda -g^{\nu \lambda} \gamma^\mu +g^{\mu \lambda} \gamma^\nu \,,$
we obtain
\begin{equation}
\label{contract1}
\cancel{D}\cancel{D}  (\gamma \cdot \psi) -D^2  (\gamma \cdot \psi)  -\cancel{D} (D \cdot \psi) + D_\mu ( \cancel{D} \psi^\mu) +m_1 \Gamma^{\mu \lambda} D_\mu \psi_\lambda +m_2 D \cdot \psi= 0 \,,
\end{equation}
where $\cancel{D} = \gamma^\mu D_\mu$.
Similarly, applying $\gamma_\mu$ to \eqref{rs1} yields
\begin{equation}
\label{contract2}
(d-2) \cancel{D}  (\gamma \cdot \psi)  +(2-d) D \cdot \psi +\left((d-1) m_1 + m_2\right)  \gamma \cdot \psi = 0 \,.
\end{equation}

Now, in exact AdS the Riemann tensor takes the simple maximally symmetric form
$R_{\rho \sigma \mu \nu} = -(g_{\rho \mu} g_{\sigma \nu} - g_{\rho \nu} g_{\sigma \mu} ) \,$, where we have set the AdS radius to one.
Using the general relation
\begin{equation}
\label{riemann}
[D_\mu, D_\nu] \psi^\lambda = R^\lambda{}_{\sigma \mu \nu} \psi^\sigma +\frac{1}{4} R_{\mu \nu \rho \sigma} \Gamma^{ \rho \sigma} \psi^\lambda
\end{equation}
combined with the above expression for the AdS Riemann tensor, we find that eq. (\ref{contract1}) becomes
\begin{equation}
\label{divergence}
D \cdot \psi = \frac{1}{m_1-m_2} \left( \frac{(d-1)(d-2)}{4} + m_1 \, \cancel{D} \right) \gamma \cdot \psi \,.
\end{equation}
Substituting this result into (\ref{contract2}) implies that $m_2 D \cdot \psi = C \gamma \cdot \psi$, where $C$ is a constant
depending on $d,m_1, m_2$.  In the case $m_2 = 0$, this relation becomes
\begin{equation}
\left(m_1^2 -\frac{(d-2)^2}{4} \right) \gamma \cdot \psi = 0 \,.
\end{equation}
Let us now define $m^2_0$ to be the special mass value given by $m_0^2 \equiv (d-2)^2/4$. Hence, if $m_1^2 \neq m^2_0$, we arrive at the gamma-traceless and divergence-free conditions
\begin{equation}
\gamma^\mu \psi_\mu = 0 = D_\mu \psi^\mu \,.\label{eqRS1}
\end{equation}
Applying these conditions to the original Rarita-Schwinger equation (\ref{rs1}) leads to the ``Dirac equation''
\begin{equation}
\label{dirac}
\left(\gamma^\mu D_\mu \ - m_1 \right) \psi_\lambda = 0 \,.
\end{equation}
Note that if $m_2 \neq 0$ or $m_1^2 = m_0^2$, we do \emph{not} obtain the divergence-free and gamma-traceless conditions from the equations of motion.
The case $m_2 = 0$, $m_1^2 = m_0^2$ deserves special attention and will be discussed in detail in section \ref{massless}. For $m_2 \neq 0$, the $\gamma$-trace of $\psi_\mu$ becomes a dynamical Dirac spinor; we will not consider such cases any further (see \cite{Koshelev:1998tu,Matlock:1999fy}). We refer to the case $m_1 \equiv m \neq m_0, m_2 = 0$ as the ``massive'' theory of the Rarita-Schwinger field.

\subsection{``Massless'' Rarita-Schwinger fields}
\label{massless}

Let us now discuss the special case $m^2 = m^2_0 = (d-2)^2/4$ and $m_2 = 0$, where it was seen above that the equations of motion do not automatically imply the gamma-traceless and divergence-free conditions. One may check that the ``massless'' spectrum of supergravity on, for example, $AdS_4 \times S^7$ \cite{Casher:1984ym} or $AdS_5 \times S^5$ \cite{Kim:1985ez} contains spin-3/2 fields with exactly these mass values. The same statement holds for $\cN = 1$ minimal AdS supergravity in $d = 4$ \cite{Townsend:1977qa}, which we discuss in section \ref{nonlinear} below.  It is therefore not surprising that for this particular value of the mass, the theory admits an extra gauge symmetry which is closely related to supersymmetry. This symmetry can be seen as follows.  First, by defining the ``superderivative''
\begin{equation}
\mathcal{D}_\nu = D_\nu -\frac{m}{d-2} \gamma_\nu \,,\label{cov_der}
\end{equation}
the Rarita-Schwinger equation can be expressed in the form $\Gamma^{\mu \nu \lambda} \mathcal{D}_\nu \psi_\lambda = 0$.  Acting on this relation with $\mathcal{D}_\mu$, we obtain the Bianchi-type identity
\begin{equation}
0 = \mathcal{D}_\mu (\Gamma^{\mu \nu \lambda} \mathcal{D}_\nu \psi_\lambda) = -\frac{(d-1)(d-2)}{4} \left(\frac{4 m^2}{(d-2)^2} -1 \right) \gamma^\sigma \psi_\sigma \,.
\end{equation}
Note that the right hand side vanishes identically in the ``massless'' case $m^2 = m_0^2$,
while in the massive case we again obtain the constraint $\gamma^\mu \psi_\mu = 0$.  As this suggests, in the massless case one can show that the action is invariant under the gauge transformation $\delta \psi_\mu = \mathcal{D}_\mu \epsilon$, where $\epsilon$ is an arbitrary spinor
 (so this is essentially a supersymmetry-type transformation).  In particular, the variation of the Rarita-Schwinger Lagrangian under these transformations is
\begin{equation}
\delta L_{RS} = \frac{N (d-2)(d-1)}{8} \left(\frac{4 m^2}{(d-2)^2} -1 \right)  \left(\overline{\epsilon} \gamma^\sigma \psi_\sigma - \overline{\psi_\sigma} \gamma^\sigma \epsilon \right),
\end{equation}
which vanishes (off-shell) in the massless case.

In fact, we can use this gauge freedom in the massless case to fix the gauge condition $\gamma^\mu \psi_\mu = 0$.  The divergence-free condition then directly follows from \eqref{divergence}.  To see this explicitly, let $\psi'_\mu = \psi_\mu +\mathcal{D}_\mu \epsilon$.  Then we require
\begin{equation}
\gamma^\mu \psi'_\mu = \gamma^\mu \psi_\mu + \left(\cancel{D} - \frac{d}{d-2}\,m \right) \epsilon = 0\,,
\end{equation}
so we get the desired result by choosing $\epsilon = - (\cancel{D}-M)^{-1} \gamma^\mu \psi_\mu$ for $M \equiv d m/(d-2)$.  Note there is still some residual gauge freedom to take $\delta \psi'_\mu = \mathcal{D}_\mu \chi$, for $\chi$ a solution of  $(\cancel{D}-M) \chi = 0$.

In summary, we have seen that by choosing an appropriate gauge in the massless case, we can write the Rarita-Schwinger equation in the same form given earlier for the massive case:  $\gamma^\mu \psi_\mu = 0, D_\mu \psi^\mu = 0, (\cancel{D}-m) \psi_\mu = 0$.  We can thus discuss in a unified way solutions to the equations of motion for both massive and massless fields.

One can also learn from the existence of an extra gauge invariance in the ``massless'' case that massive fields ($m_1 \neq (d-2)/2$, $m_2$ arbitrary) should admit an extra spin-1/2 degree of freedom\footnote{We thank Don Marolf for discussions of this point.}. However, as further explained in section \ref{sec:sol}, this extra spinorial degree of freedom will have no significant role in our analysis as long as $ 2m_1 < d$.

\subsection{General asymptotic solutions}
\label{sec:sol}

In order to describe the boundary fields and determine the allowed boundary conditions, we must first analyze the behavior of solutions to \eqref{eqRS1}, \eqref{dirac} near the AdS boundary. Evaluating the on-shell renormalized action also requires only the asymptotic solutions, since the Rarita-Schwinger action is zero on-shell (see section \ref{renorm}). Exact solutions to the Rarita-Schwinger equation have been found by Fourier-transforming to momentum space \cite{Deser:1984py,Volovich:1998tj,Koshelev:1998tu,Corley:1998qg,Rashkov:1999ji,Matlock:1999fy,Rychkov:1999zp}, but for our purposes it is useful to solve the equations by performing a Fefferman-Graham type expansion around the boundary.

To do this expansion,  it will be convenient to use the conformal compactification of AdS spacetime.  In Poincar\'e coordinates, the AdS metric takes the form
\begin{equation}
\label{metricd} ds^2 = \frac{1}{\Omega^2} \left(- dt^2 + d\Omega^2 + dx_1^2+\ldots dx_{d-2}^2 \right) \,.
\end{equation}
  One then defines
an unphysical metric $\tilde g_{\mu \nu} = \Omega^2 g_{\mu \nu}$ so that the unphysical spacetime is a manifold with boundary $\I \cong
\mathbb{R}^{d-1}$ at $\Omega = 0$.  In this spacetime, $\tilde n_\mu = -\tilde D_\mu \Omega$ coincides
with the unit normal to the boundary, where $\tilde D_\mu$ is the
torsion-free covariant derivative compatible with $\tilde g_{\mu \nu}$.
It is also useful to define the orthogonal projector $\thh_{\mu \nu} =
\tg_{\mu \nu} - \tn_\mu \tn_\nu$, which at $\Omega =0$ becomes the induced
metric on the boundary
$\thh_{\mu \nu} dx^\mu dx^\nu \,|_\I = -dt^2+ dx_1^2+\ldots dx_{d-2}^2$,
i.e., Minkowski space.  Indices on all tensor fields
with a tilde are raised and lowered with the unphysical metric
$\tg_{\mu \nu}$ and its inverse $\tg^{\mu \nu}$.  Note that because we work in the Poincar\'e patch, $\tgam_\mu \equiv \Omega \gamma_\mu$ are now flat-space gamma matrices and we may identify $ \tD_\mu \to \partial_\mu$.

Under the conformal transformation $\tg_{\mu \nu} = \Omega^2 g_{\mu \nu}$, the equation of motion (\ref{dirac}) takes the form
\begin{equation}
\label{condirac}
\Omega \tgam^\rho \tD_\rho \psi_\sigma +\frac{d-3}{2} \tn_\rho \tgam^\rho \psi_\sigma - m\psi_\sigma
+\tgam_\sigma \tn^\lambda \psi_\lambda = 0 \,.
\end{equation}
By solving the equation \eqref{condirac} at leading order in $\Omega$, we are lead to the existence of two independent solutions with fall-off $O(\Omega^{\frac{d-3}{2} \pm m})$. We now distinguish between two general cases: $2m$ integer valued and $2m$ non-integer valued.  Without loss of generality, we take $m \geq 0$ in the following.

\begin{enumerate}

\item Case: $2m \not \in \mathbb{Z}$

Let $k$ be the greatest integer strictly less than $2m$, i.e., $0< 2m-k < 1$.  We then expand the Rarita-Schwinger field as
\begin{equation}
\psi_\mu = \Omega^{\frac{d-3}{2}-m} \sum^{k}_{n = 0} \alpha^{(n)}_\mu \Omega^n  +\beta^{(0)}_\mu \Omega^{\frac{d-3}{2}+m} + O(\Omega^{\frac{d-1}{2}+k-m}) \,,\label{exp1}
\end{equation}
where the coefficients $\alpha^{(n)}_\mu, \beta^{(n)}_\mu$ are vector-spinors that do not depend on $\Omega$, but may in general depend on time and the remaining spatial coordinates.  Substituting this expansion into the equations of motion gives certain constraints on the coefficients.  The gamma-traceless constraint $\gamma^\mu \psi_\mu = 0$ clearly implies that all the coefficients are themselves gamma-traceless: $\tgam^\mu \alpha^{(j)}_\mu = 0 =\tgam^\mu \beta^{(j)}_\mu$.
For terms up to $O(\Omega^{\frac{d-3}{2}+m})$, we find from the remaining equations of motion
\begin{equation}
\tn^\mu \alpha^{(0)}_\mu = 0, \quad \tP_- \alpha^{(0)}_\mu  = 0, \quad \tn^\mu \beta^{(0)}_\mu = 0, \quad  \tP_+ \beta^{(0)}_\lambda  =0\,,
\end{equation}
and for $k = 1, \ldots, n$
\begin{equation}
\tn^\mu \alpha^{(k)}_\mu = - \frac{2}{d-2k +2m} \, \thh^{\mu \nu} \partial_\mu \alpha^{(k-1)}_\nu
\end{equation}
\begin{equation}
\thh_{\mu}{}^\lambda \alpha^{(k)}_\lambda  = \frac{\left(m+(-1)^{k}(m-k)\right)}{k(2m-k)} \, \thh_{\mu}{}^\lambda \thh^{\rho \sigma} \left[ \tgam_\rho \partial_\sigma \alpha^{(k-1)}_\lambda -\frac{2}{d-2k+2m} \, \tgam_\lambda \partial_\rho \alpha^{(k-1)}_\sigma \right] \,,
\end{equation}
where we have defined the radial gamma matrix projectors $\tP_\pm = \frac{1}{2} (1 \pm \tn_\mu \tgam^\mu)$. We see that the solution is parameterized by the two vector-spinor boundary fields $\alpha^{(0)}_{i,+} \equiv \tP_+ \alpha^{(0)}_i$ and $\beta^{(0)}_{i,-} \equiv \tP_- \tilde \beta^{(0)}_i$.  Here we have introduced the useful index notation $\mu = (\Omega, i)$, i.e. $i,j,k\ldots$ denote indices tangent to the boundary. These indices are then raised and lowered with the Minkowski metric $\eta_{ij}$ and its inverse. The coefficients in the radial expansion at order $O(\Omega^{\frac{d-1}{2} - m+k})$ and beyond are completely determined in terms of these two independent vector-spinors and will not play any role in the analysis below.

\item Case: $2m \in \mathbb{Z}$

The second general class of solutions occurs when $m$ is an integer or half-integer. Note that the massless case always lies in this category.  Here, the spin-3/2 field can be expanded as
\begin{equation}
\psi_\mu = \Omega^{\frac{d-3}{2}-m} \sum^{2m-1}_{n = 0} \alpha^{(n)}_\mu \Omega^n + \alpha^{(2m)}_\mu \Omega^{\frac{d-3}{2}+m} \log \Omega +\beta^{(0)}_\mu \Omega^{\frac{d-3}{2}+m} + \ldots \,. \label{exp}
\end{equation}
The coefficients are once again all gamma-traceless and now satisfy
\begin{equation}
\tn^\mu \alpha^{(0)}_\mu = 0, \quad m \tP_- \alpha^{(0)}_\mu  = 0,
\end{equation}
\begin{equation}
\left(\frac{d}{2} - m\right) \tn^\mu \beta^{(0)}_\mu = -\thh^{\mu \nu} \partial_\mu \alpha^{(2m-1)}_\nu, \quad  m \tP_+ \beta^{(0)}_\lambda  = \frac{1}{2} \tP_+ \left(\thh^{\rho \sigma} \tgam_\sigma \partial_\rho \alpha^{(2m -1)}_\mu + \tgam_\mu \tn^\rho \beta^{(0)}_\rho \right)\,,
\end{equation}
while for $k = 1, \ldots, 2m-1$
\begin{equation}
\left(\frac{d}{2}+m -k\right) \tn^\mu \alpha^{(k)}_\mu = -\thh^{\mu \nu} \partial_\mu \alpha^{(k-1)}_\nu \label{eq:20}
\end{equation}
and
\begin{equation}
\alpha^{(k)}_\mu  = \frac{\left(m+(m-k)\tn^\rho \tgam_\rho \right)}{k(2m-k)} \, \left[ \thh^{\rho \sigma} \tgam_\rho \partial_\sigma \alpha^{(k-1)}_\mu + \tgam_\mu \tn^\rho \alpha^{(k)}_\rho  \right] \,.\label{eq:18}
\end{equation}
The coefficient of the logarithmic term $\alpha^{(2m)}_\mu$ is only nonzero when $m \in \mathbb{Z}+\frac{1}{2}$, and in this case
\begin{equation}
\left(\frac{d}{2} - m\right) \tn^\mu \alpha^{(2m)}_\mu = 0,
\end{equation}
\begin{equation}
 \quad  \tP_+ \alpha^{(2m)}_\mu = 0 \,, \quad
\tP_- \alpha^{(2m)}_\mu  = -\tP_- \left(\thh^{\rho \sigma} \tgam_\sigma \partial_\rho \alpha^{(2m -1)}_\mu + \tgam_\mu \tn^\rho \beta^{(0)}_\rho \right) \,.
\end{equation}
When $m = 0$, the linearly independent modes degenerate to a single unconstrained boundary field $\beta^{(0)}_i$, which we then naturally
split with $\tP_\pm$ into two independent parts that we call $\alpha^{(0)}_{i,+}$, $\beta^{(0)}_{i,-}$.

It is straightforward to check that the right-hand side of \eqref{eq:20} is also proportional to $(\frac d 2 + m - k)$ for $k\geq 2$ as a consequence of \eqref{eq:18}. Thus, the above relations indicate that for certain values of the mass and spacetime dimension (i.e., when this factor vanishes), the radial components of certain coefficients are undetermined by the equations of motion. This reflects the existence of an extra spin-1/2 degree of freedom which admits the expansion \eqref{exp} when $m$ is half-integer or integer valued. This degree of freedom was not seen in the first class of solutions because it falls off as $O(\Omega^{d-3/2})$ and the ansatz \eqref{exp1} did not include integer or half-integer powers of $\Omega$. The extra degree of freedom appears in the coefficients $\tilde n^\mu \alpha_\mu^{(2m)}$, $\tilde n^\mu \beta_{\mu}^{(0)}$ for $m = \frac d 2$ and in the coefficient $\tilde n^\mu \alpha_\mu^{(m+\frac d 2)}$ when $m = \frac d 2 + j$, $j \geq 1$. In the following discussion of the on-shell action, the counterterms, and the normalizeability, it turns out that one only needs to solve the equations of motion up to order $\Omega^{\frac{d-3}{2}+m}$. Hence, this phenomenon will only effect the counterterms if $m \geq \frac d 2$; we will not consider such cases below (note that because $m_0 < \frac d 2$, this restriction does not exclude the interesting ``massless'' cases).

The occurrence of logarithmic modes for half integer masses hints at the appearance of conformal anomalies that will be described below.

\end{enumerate}

\subsection{Standard normalizeability}
\label{standard}

In this section, we analyze normalizeability of the above solutions with respect to the standard symplectic structure. Standard methods applied to the action \eqref{rsaction} lead to
\begin{equation}
(\omega)_{\mu_1 \ldots \mu_{d-1}} = N \left(\overline{\delta_1 \psi_\mu}  \Gamma^{\mu \nu \lambda} \delta_2 \psi_\lambda -
\overline{\delta_2 \psi_\mu}  \Gamma^{\mu \nu \lambda} \delta_1 \psi_\lambda \right) \epsilon_{\nu \mu_1 \ldots \mu_{d-1}} \,.
\end{equation}
This expression simplifies after using the constraint $\gamma^\mu \psi_\mu = 0$, resulting in the symplectic structure
\begin{equation}
\label{RSstructure}
\sigma_{\Sigma} (\delta_1\psi_\mu, \delta_2 \psi_\nu)  = N \int_\Sigma d^{\,d-1}x \sqrt{g_\Sigma} \, t_\nu
\left(\overline{\delta_1 \psi^\mu}  \gamma^\nu \delta_2 \psi_\mu - \overline{\delta_2 \psi^\mu}  \gamma^\nu \delta_1 \psi_\mu\right) \,.
\end{equation}
Substituting the asymptotic form of the slow fall-off mode (i.e. $\psi_\mu \sim O(\Omega^{\frac{d-3}{2} - |m|}))$  into the symplectic structure, we find that the integrand near $\Omega \to 0$ is $O(\Omega^{-|2m|})$. Therefore, the standard symplectic structure for the slow fall-off mode is finite only for fields in the mass range $|m| < 1/2$, where there is then a choice of boundary conditions.  This mass range is thus analogous to the range at or slightly above the Breitenlohner-Freedman bound for scalar fields \cite{Breitenlohner:1982jf}. However,  normalizeability requires boundary conditions of Dirichlet-type for the limiting case $m = 1/2$, where a logarithmic mode appears. The corresponding results for spin-1/2 fields are essentially identical  \cite{Amsel:2008iz}.  It was further argued in \cite{Amsel:2008iz} that for real values of the spin-1/2 fermion mass,  there is no analogue of unstable modes below the Breitenlohner-Freedman bound.  We would similarly expect that there are no such unstable modes in the spin-3/2 case, though this has yet to be checked explicitly.

The massless spin-3/2 fields are notably outside this special mass range. Indeed, for all $d \geq 3$, we have $|m_0| \geq 1/2$. Thus, taking $m = m_0 >0$ and using the standard choice for the symplectic structure, we are required to choose the ``Dirichlet'' boundary condition that fixes the slow fall-off mode to zero. This implies that the behavior near the boundary is $\psi_\mu  = O(\Omega^{d-\frac{5}{2}})$, which matches the boundary conditions imposed for $d = 4$ in \cite{Hollands:2006zu}.

Normalizeability places no restriction on the boundary fields, but
we must still impose conservation of the symplectic structure.  Using the expansion (\ref{exp1}), we find that for $|m|< 1/2$, the symplectic flux through the boundary is given by
\begin{equation}
\F = \int_\I \boldsymbol{\omega} = -N \int_\I d^{\,d-1}x \,  \left[ \left(\overline{ \delta_1 \alpha^{(0)}_{i,+}} \delta_2 \beta^{(0),i}_- - \overline{ \delta_1 \beta^{(0)}_{i,-}} \delta_2 \alpha^{(0),i}_+ \right) - \Big(\delta_1 \leftrightarrow \delta_2 \Big) \right] \,.\label{fluxsol}
\end{equation}
Any boundary condition such that the integrand vanishes is sufficient to give well defined dynamics.
For example, a non-trivial mixed condition in even dimensions is given by the linear relation $\beta^{(0)}_{i,-} = i q \gamma_{d+1} \alpha^{(0)}_{i,+}$, where $q$ is an arbitrary real parameter and $\gamma_{d+1}$ is the higher dimensional generalization of $\gamma_5$ in four dimensions.  Dirichlet or Neumann boundary conditions correspond to the particular choices $q = \infty$ or $q = 0$, respectively.

\subsection{Renormalized modes}
\label{renorm}

We next turn to the case $m \geq 1/2$. For these masses, the mode $\alpha^{(0)}$ is non-normalizeable with respect to the standard symplectic structure. However, as in the case of gravitational perturbations \cite{Compere:2008us}, one can consider the renormalized symplectic structure \eqref{renomega} for which we expect the slow fall-off modes to be normalizeable as well, allowing more general boundary conditions. In what follows, we will determine the boundary counterterms required to renormalize the symplectic structure associated with the action \eqref{rsaction}. We then verify normalizeability of the slow fall-off modes for a large class of parameters.

For convenience, we will assume below that the spin-3/2 fields obey the Majorana ``reality'' condition, with the charge conjugation matrix satisfying the same properties as in $d = 4$ (see appendix \ref{sec:conventions}).  Hence, our results will be strictly valid in $d = 2, 3, 4 \mod 8$ dimensions, but could be easily generalized to other dimensions by introducing extra signs in certain places or working with complex Dirac spinors.

In order to determine how to modify the symplectic structure, we must first discuss boundary counterterms in the action:
\begin{equation}
S =  S_{RS} + \int_\I d^{d-1}x \, L_{ct} \,,\label{actionRStot}
\end{equation}
where $S_{RS}$ is the standard Rarita-Schwinger action and the boundary counterterm Lagrangian $L_{ct}$ is to be determined.  Note that because the Rarita-Schwinger action vanishes on-shell, $S$ will be manifestly finite on-shell if we also take $ S_{ct}$ finite.  In other words, $S_{ct}$ is not a divergent counterterm, but is simply the finite expression required to provide a valid variational principle.
Using the asymptotic solutions of $\psi_\mu$ given in \eqref{exp1} and \eqref{exp}, we find in general that on-shell
\begin{eqnarray}
\delta S_{RS} &=& -N\int_\I d^{\,d-1}x \left[ \left( \rm{divergent\,terms}\right) +   \overline{\alpha^{(0)}_{i,+}} \delta \beta^{(0),i}_- - \overline{\beta^{(0),i}_-} \delta \alpha^{(0)}_{i,+} +\overline{F^i[\alpha_+^{(0)}]}\delta \alpha^{(0)}_{i,+} \right] \,\label{eq:19}
\end{eqnarray}
where $F^i[\alpha_+^{(0)}]$ refers to possible additional finite terms involving at least 3 derivatives of $\alpha^{(0)}_+$ that may only arise for $m \geq 3/2$ and $m \in \mathbb Z +\frac{1}{2}$.  By the argument above, the divergent terms must combine to form a total derivative and can thus be dropped.  We have checked that this is true explicitly for all $|m| < 3/2$ in any $d\geq 3$ and $|m| = 3/2$ in any $d>3$.

To impose Dirichlet-type boundary conditions (fixing $\alpha^{(0)}_{i, +}$), one has to add to the original action a  counterterm of the form
\begin{equation}
L_{ct} =  N \, \overline{\alpha^{(0)}_{i, +}}  \beta_{-}^{(0),i} + L_{fin}[\alpha^{(0)}_{+}] \,, \label{count}
\end{equation}
where $L_{fin}[\alpha_+^{(0)}]$ are non-minimal terms depending  on $\alpha^{(0)}_{+}$ and its derivatives.  The total action then obeys
\begin{equation}
\delta S =  \delta S_{RS} + \delta \int_\cI d^{\,d-1}x L_{ct} =  \int_\I d^{\,d-1}x \;  \overline{\pi^{(0),i}_-} \delta \alpha^{(0)}_{i,+}  \,.
\end{equation}
Here, the conjugate field is found to be $\overline{\pi^{(0),i}_-} \equiv 2 N  \overline{\beta^{(0),i}_-}-N \overline{F^i[\alpha^{(0)}_+]} + \varQ{L_{fin}}{\alpha^{(0)}_{i,+}} $. The symplectic flux has the general form \eqref{flux}. Neumann-type boundary conditions (fixing $\pi^{(0),i}_-$) are then obtained by performing the Legendre transformation $S_{Neu} = S - \int_\I \overline{\pi^{(0),i}_-}\alpha^{(0)}_{i,+}$.

We now make the observation that the first term in the right-hand side of \eqref{count} can be expressed in terms of the original bulk fields as
\begin{eqnarray}
\overline{\alpha^{(0)}_{i, +}}  \beta^{(0),i}_- =
     \frac{1}{2} \sqrt{-h} \, h^{i j} \overline{\psi_i} \psi_j + \text{(divergent terms)} + G_{fin}[\alpha^{(0)}_+] \label{count2}
\end{eqnarray}
where the number of divergent terms (depending only on $\alpha^{(0)}_+$ and its derivatives) increases with the value of the mass.  The various finite terms $G_{fin}[\alpha^{(0)}_+] $ (depending also only on $\alpha^{(0)}_+$ and its derivatives) appear for masses $m \geq 3/2, m \in \mathbb Z +\frac{1}{2}$. We conclude that the counterterm of the form \eqref{count} containing the minimal number of terms when expressed in terms of the bulk field $\psi_\mu$ is given by the first term on the right-hand side of \eqref{count2} plus a finite number of terms canceling the additional divergences. The resulting minimal counterterm Lagrangian $L_{ct}$ for $m \leq 3/2$ is found to be
\begin{eqnarray}
L_{ct} &=&   \frac{N}{2} \sqrt{-h} \Bigg[\overline{\psi_i} \psi^i + 2 f_{1/2}(\Omega)  \overline{\psi_i} \gamma^j \partial_j \psi^i  \nonumber\\
&& \quad + 2 f_{3/2}(\Omega) \left( \overline{\psi_i} \gamma^j \partial^k \partial_k \partial_j \psi^i
- \frac{4 d}{(d-1)(d+1)} \overline{\psi^i} \gamma^j \partial_i \partial_j \partial_k \psi^k \right)  \Bigg] \,,\label{fermcounter}
\end{eqnarray}
where
\begin{equation}
f_{1/2}(\Omega) = \left\{ \begin{array}{ll}
\frac{1}{1-2m} & \textrm{if $m \neq \frac{1}{2}$}\\
\log \Omega & \textrm{if $m = \frac{1}{2}$}
\end{array} \right. \,\qquad
f_{3/2}(\Omega) = \left\{ \begin{array}{ll}
0 & \textrm{if $m < \frac{3}{2}$}\\
\frac{1}{4} \log \Omega & \textrm{if $m = \frac{3}{2}$,\;}d>3,\\
\end{array} \right.
\end{equation}
and where all indices are raised with $h^{ij} = \Omega^2 \eta^{ij}$. For $3/2 < m < 5/2$, we expect the appearance of additional counterterms involving 3 derivatives, while for $m \geq 5/2$ we would require additional counterterms involving more than 3 derivatives. The properties of the recursion relations for the coefficients $\alpha^{(n)}$ found in section \ref{sec:sol} imply that all such counterterms will involve an odd number of derivatives. This is consistent with the fact that acting with supersymmetry transformations on gravity counterterms \cite{Balasubramanian:1999re}  generates only fermionic counterterms with an odd number of derivatives.

For half-integer masses, we have seen that logarithmic modes always arise in the asymptotic solutions to the Rarita-Schwinger equation.  Furthermore, we have found in the specific cases $m = 1/2, 3/2$ that the required counterterms contain explicit $\log \Omega$ dependence, and we would expect this to be true for larger half-integer masses as well.  This suggests that for all $m \in \mathbb Z +\frac{1}{2}$, the renormalization procedure explicitly breaks radial diffeomorphism invariance.  This is interpreted in the AdS/CFT language as a dynamical breaking of conformal invariance \cite{Henningson:1998gx}. This phenomenon is partially explained by the analogous results in gravity. Indeed, the $m = 1/2$ field appears in $d=3$  supergravity, as does $m = 3/2$ for $d = 5$. In these theories, the gravitational counterterms contain a logarithmic term (proportional to the Einstein-Hilbert action for $d=3$ or  Weyl curvature squared for $d = 5$) that breaks conformal invariance. We point out though that the spin-3/2 conformal symmetry breaking is generic for any half-integer $m$ in \emph{any} dimension.

We end this section by providing a non-trivial check of the claim that the renormalized symplectic structure is finite for the ``non-normalizeable'' mode and by providing a non-trivial mixed boundary condition. Let us consider in particular the cases $1/2 < m \leq 1$, $d \geq 3$. Following the general discussion in section \ref{prelim}, we can vary the counterterm action (for the Neumann theory) as $\delta {\bf L}_{Neu} = \frac{\delta {\bf L}_{Neu}}{\delta \psi_i} \delta \psi_i + d \boldsymbol{\theta}_{ct}$ and obtain $d \boldsymbol{\theta}_{ct} =  d^{d-1}x \sqrt{-h} \,\partial_j \left(\frac{N}{2m-1} \overline{\psi_i} \gamma^j \delta \psi^i \right) $. Using the expansion \eqref{exp}, we find
\begin{equation}
(\omega_{ct})_{i_1 \ldots i_{d-2}} |_\I = \frac{2N}{2m-1} \,\Omega^{1-2m} \overline{\delta_1 \alpha^{(0)}_{k,+}} \tgam^l \delta_2 \alpha^{(0),k}_+ \, \tilde \epsilon_{\Omega l i_1 \ldots  i_{d-2}}\,.
 \end{equation}
This divergent term is then recognized as the exact divergent term in the standard symplectic structure.  The renormalized symplectic structure \eqref{renormstructure} takes the form
\begin{equation}
\sigma_{\Sigma, ren} = 2N \int_\Sigma d^{\,d-2} x d\Omega \left[\Omega^{3-d} \tg^{\mu \lambda} \overline{\delta_1 \psi_\mu} \tgam^t \delta_2 \psi_\lambda -\Omega^{-2m} \thh^{\mu \lambda}  \overline{\delta_1 \alpha^{(0)}_\mu} \tgam^t \delta_2 \alpha^{(0)}_\lambda \right] \,.\label{renomex}
\end{equation}
The first term in the above expression diverges as $O(\Omega^{-2 m})$, but  this divergence is exactly canceled by the second term. The remaining terms are then of order $O(\Omega^{0})$ and so $\sigma_{\Sigma, ren}$ is manifestly finite.  After the divergences are removed, the renormalized symplectic flux takes the same (finite) form as in \eqref{fluxsol}.  This explicit check of our methods includes in particular the special case $m = 1$, $d = 4$, which is relevant for $\cN = 1$ minimal AdS$_4$ supergravity.  To have a mixed boundary condition in this case, we can, for example, impose the linear relation $\beta^{(0)}_{i,-} = i q  \gamma_5 \alpha^{(0)}_{i,+} $ which leads to vanishing flux and includes the Neumann boundary conditions as the case $q = 0$.  This corresponds to a deformation of the Neumann theory given by
\begin{eqnarray}
 \int_\I d^{\, 3}x \sqrt{-h} L_{ct,q} =
     -\frac{N}{2}  \int_\I d^{\,3}x \sqrt{-h} \, \left(\overline{\psi_i} \psi^i - 2 \overline{\psi_i} \gamma^j \partial_j \psi^i  -2 i q \Omega^2 \overline{\psi_i} \gamma_5 \psi^i \right) \,.
\end{eqnarray}
For $q \neq 0$, these mixed boundary conditions break conformal invariance, as can also be seen in the explicit radial dependence of the counterterm.


\subsection{Euclidean boundary propagator}
\label{prop}

In this section, we review the two-point correlation function of the CFT operator dual to the boundary field $\alpha^{(0)}_{i,+}$. This correlation function can be obtained from the on-shell action and the asymptotic solutions derived in the previous section via standard methods \cite{Volovich:1998tj,Koshelev:1998tu,Corley:1998qg,Rashkov:1999ji,Matlock:1999fy,Rychkov:1999zp}. We then discuss the correlation function for spin-3/2 fields with Neumann boundary conditions, following the analysis of \cite{Klebanov:1999tb} performed for scalar fields in the Breitenlohner-Freedman mass range.

The AdS/CFT dictionary states that the (renormalized) Euclidean Rarita-Schwinger action (viewed as a functional of the boundary fields $\alpha^{(0)}_{i,+}$) is the generating functional for correlation functions of the CFT operator dual to $\alpha^{(0)}_{i,+}$, which is the supersymmetry current. The field $\pi_{(0),-}^{i} = \beta_{(0),-}^{i} + \dots$ conjugate to $\alpha^{(0)}_{i,+}$ is interpreted as the vacuum expectation value of the dual CFT operator and is determined in the Euclidean setting in terms of $\alpha^{(0)}_{i,+}$ by imposing regularity of the solution. The two-point function of the supersymmetry current can then be obtained by substituting the asymptotic solutions \eqref{exp1} and \eqref{exp} into the on-shell Euclidean renormalized action \eqref{actionRStot}. The result is  \cite{Volovich:1998tj,Koshelev:1998tu,Corley:1998qg,Rashkov:1999ji,Matlock:1999fy,Rychkov:1999zp}
\begin{eqnarray}
S \sim \int_\cI d^{d-1} x \int_\cI d^{d-1} y \, \overline{\alpha^{(0)}_{i,+}}(x) \frac{\gamma^k (x-y)_k}{|x-y|^{2\Delta_++1}} \left( \delta^{ij} - 2 \frac{(x-y)^i (x-y)^j}{|x-y|^2} \right) \alpha^{(0)}_{j,+}(y) \,,\label{kin}
\end{eqnarray}
where $\delta_{ij}$ is the Euclidean continuation of the boundary metric $\tilde h_{ij}|_{\cI}$. The dimension of the dual supersymmetry current is given by $\Delta_+ = \frac{d-1}{2}+m$.

Now, we have seen that for any $m$ and $d$, we can perform a Legendre transformation and consider the Neumann problem, where $\pi_{(0),-}^{i}$ is fixed.  The roles of $\alpha^{(0)}_{i,+}$, $\pi_{(0),-}^{i}$ are thus interchanged, and the two-point correlation function is related to the Dirichlet result by a Legendre transformation. Following \cite{Klebanov:1999tb}, we first define $J_{Neu}[\alpha^{(0)}_{+},\pi_{(0),-}] = S[\alpha^{(0)}_{+}] - \int_\I \overline{\pi^{(0),i}_-}\alpha^{(0)}_{i,+}$. The Legendre transformed functional $L_{Neu}[\pi_{(0),-}]$ is then obtained by minimizing $J_{Neu}[\alpha^{(0)}_{+},\pi_{(0),-}]$ with respect to $\alpha^{(0)}_{i,+}$. In order to compute $S_{Neu}$, we first Fourier transform the two-point function \eqref{kin}, which yields $S \sim \int \frac{d^d k}{(2\pi)^d} \alpha^{(0)}_{i,+}(k)  O^{ij}(k) \alpha^{(0)}_{i,+}(-k)$ with
\begin{eqnarray}
 O^{ij}(k) &=& \Pi^{il}\frac{k_k \gamma^k}{k^{1-2m}}\left( \delta_{lm} +C(m,d)\frac{k_l k_m}{k^2} \right) \Pi^{mj}, \quad m \neq \frac 1 2 + \mathbb Z,\label{eq:propRS1i}\\
&=& \log \left({\frac k \mu}\right) \Pi^{il}\frac{k_k \gamma^k}{k^{1-2m}}\left( \delta_{lm} +C(m,d) \frac{k_l k_m}{k^2} \right) \Pi^{mj},\,\,  m = \frac 1 2 + \mathbb Z \,.\label{eq:propRSi}
\end{eqnarray}
Here $k = \sqrt{k^i k_i}$, $\Pi_{i}^j = \delta_{i}^j-\frac{1}{d-1}\gamma_i \gamma^j$ is the gamma-traceless projector, $\mu$ is an additional scale appearing in the Fourier transformation when $m$ is half-integer, and $C(m,d)$ are coefficients whose precise value will not be needed for our arguments. The inverse of $ O^{ij}(k)$, defined as $ O^{jk}(k) O^{-1}_{ij}(k)=\Pi_i^k $, is given by
\begin{eqnarray}
 O^{-1}_{ij}(k) &=& \Pi_{il} \frac{k_k \gamma^k}{k^{2m+1}}\left( \delta^{lm} +C^\prime (m,d)\frac{k^l k^m}{k^2} \right) \Pi_{mj}, \quad m \neq \frac 1 2 + \mathbb Z,\label{eq:propRS1}\\
&=& \Pi_{il} \frac{k_k \gamma^k}{k^{2m+1}\log{\frac{k}{\mu}}}\left( \delta^{lm}+C^\prime (m,d) \frac{k^l k^m}{k^2} \right) \Pi_{mj} , \quad m = \frac 1 2 + \mathbb Z \label{eq:propRS} \, .
\end{eqnarray}
where $C^\prime (m,d)$ are different coefficients. The Legendre transformed functional $L_{Neu}[\pi_{(0),-}]$ is then  $L_{Neu}[\pi_{(0),-}] \sim \int \frac{d^d k}{(2\pi)^d} \pi_{(0),-}^{i}(k)  O^{-1}_{ij}(k) \pi_{(0),-}^{j}(-k) $. Given the scaling dimension of a CFT operator, the two-point function is entirely determined by conformal invariance. The scaling dimension of the dual operator in the Neumann theory with $ m \neq \frac 1 2 + \mathbb Z$ can be deduced by noticing that the prefactor in $O^{-1}_{ij}(k)$ has the same form as the one in $O^{ij}(k)$ with $m \rightarrow -m$. Therefore, $\Delta_- \equiv \frac{d-1}{2} - m$. The two-point function is given by Fourier transforming back to coordinate space,
\begin{eqnarray}
S_{Neu} \sim \int_\cI d^{d-1} x \int_\cI d^{d-1} y \, \overline{ \pi_{(0),-}^{i}}(x) \frac{\gamma^k (x-y)_k}{|x-y|^{2\Delta_- +1}} \left( \delta_{ij} - 2 \frac{(x-y)_i (x-y)_j}{|x-y|^2} \right) \pi^j_{(0),-}(y) \,.  \label{kinapp}
\end{eqnarray}
We therefore obtain a result similar to that for scalar fields \cite{Klebanov:1999tb}: the Dirichlet and Neumann theories are associated with dual CFT operators whose dimensions are $\Delta_+$ and $\Delta_-$, respectively. For $m = \frac 1 2 + \mathbb Z $, conformal invariance is broken and we were not able to Fourier transform back to coordinate space. Furthermore, the propagator \eqref{eq:propRS} has an additional pole at $k = \mu$, which shows that the Lorentzian propagator admits a tachyon. In odd spacetime dimensions $d$ such that $m = \frac{d-2}{2}$, this is natural in view of the results in gravity (see e.g. \cite{Compere:2008us}).  When the boundary is even-dimensional, there is an anomaly breaking conformal invariance and the graviton propagator admits a tachyon. For Rarita-Schwinger fields, we see that these two effects appear in any dimension as long as the mass is half-integer.  The exact propagators \eqref{eq:propRS1}-\eqref{eq:propRS} have a branch cut either because of the square root in $k$ or because of the logarithm. This is similar to the behavior of the boundary graviton described in \cite{Tomboulis:1977jk,Gubser:1998bc,Hawking:2000bb,Compere:2008us}. It implies that the propagator contains a continuous spectrum of bound states with masses ranging from zero to infinity.

\section{Boundary analysis of $d=4$, $\cN=1$ AdS supergravity}
\label{nonlinear}

In this section, we analyze the non-linear theory of a graviton multiplet in an asymptotically AdS spacetime and classify boundary conditions that preserve supersymmetry.  We begin by specializing to four dimensions and coupling the Rarita-Schwinger field non-minimally (due to torsion terms) to gravity to obtain $\cN = 1$ supergravity.  We then mainly follow the holographic renormalization procedure  \cite{deHaro:2000xn,Bianchi:2001kw,Bianchi:2001de}.  After discussing the general asymptotic solutions in Fefferman-Graham form, we determine the set of gauge transformations that preserve the boundary conditions.  By renormalizing the action, we find the boundary stress-tensor and supercurrent dual to the boundary vierbein and Rarita-Schwinger field.  We then show that the renormalized action is invariant (non-anomalous) under all gauge transformations at the boundary. We conclude by discussing Neumann boundary conditions for the boundary supergravity multiplet, as well as certain deformations of the Neumann theory.

\subsection{Preliminaries}

The $\cN=1$, $d=4$ supergravity action with cosmological constant $\Lambda = -3$ is given by \cite{Deser:1976eh,Townsend:1977qa}
\begin{equation}
\mathcal L_{sugra}=\frac{1}{2 \kappa^2} e \,(R-2\Lambda) - \frac{e}{2} \left( \bar \psi_\mu \Gamma^{\mu\rho\sigma}D_\rho(\omega) \psi_\sigma +\bar \psi_\mu \Gamma^{\mu \sigma}\psi_\sigma\right) \label{sugraaction}
\end{equation}
(see also the review in \cite{vanNieuwenhuizen:2006pz}).
It will be convenient to express this action in the  alternative form
\begin{equation}
\mathcal L_{sugra}= -\frac{1}{8 \kappa^2} \varepsilon^{\mu\nu\rho\sigma} \epsilon_{\hat \mu\hat \nu\hat \rho\hat\sigma} R^{\;\;\;\hat\mu\hat\nu}_{\mu\nu}(\omega) e_\rho^{\hat\rho} e_\sigma^{\hat \sigma} - \frac \Lambda {\kappa^2} e - \frac{i}{2}\varepsilon^{\mu\nu\rho\sigma} \bar \psi_\mu \gamma_5 \gamma_\nu \mathcal D_\rho(\omega) \psi_\sigma .
\end{equation}
The equivalence of the two Lagrangians easily follows as a consequence of the properties $\varepsilon^{\mu\nu\rho\sigma} \epsilon_{\hat \mu\hat \nu\hat \rho\hat\sigma} e^{\hat \rho}_\rho e^{\hat\sigma}_\sigma= -4 e e^\mu_{[\hat\mu}e^\nu_{\hat\nu]}$ and $\Gamma^{\hat\mu\hat\nu\hat\rho}=i \epsilon^{\hat \mu\hat \nu\hat \rho\hat\sigma}\gamma_5 \gamma_{\hat\sigma}$. The covariant derivative is $\mathcal D_\mu(\omega) = D_\mu(\omega) - \frac 1 2 \gamma_\mu$ and $D_\mu \psi_\nu$ is given by \eqref{dpsi} but with $\omega(e)$ replaced by $\omega = \omega(e, \psi)$, which includes torsion terms. We will use the 1.5 formalism for supergravity, which is similar to $2^{\rm nd}$ order formalism except that one does not apply the chain rule to $\omega$ when varying the action (see \cite{vanNieuwenhuizen:2006pz}). The spin connection is defined as a solution of the constraint
\begin{equation}
D_{[\mu}(\omega) e_{\nu]}^{\hat \mu} = \frac{\kappa^2}{4}\bar \psi_\mu \gamma^{\hat \mu} \psi_\nu ,\label{eq:spin}
\end{equation}
which is the analogue of the ``vierbein postulate'' discussed in the linearized theory of section \ref{massive}.  The solution takes the form $\omega_{\mu \hat\mu\hat\nu} = \omega_{\mu \hat\mu\hat\nu}(e)+\kappa_{\mu \hat\mu\hat\nu}$, where $\omega_{\mu \hat\mu\hat\nu}(e)$ is the textbook connection given in terms of the vierbein (see \eqref{textbookomega}) and $\kappa_{\mu \hat\mu\hat\nu} = \frac{\kappa^2}{4}(\bar \psi_\mu \gamma_{\hat \mu} \psi_{\hat \nu} - \bar \psi_\mu \gamma_{\hat \nu} \psi_{\hat \mu} +\bar \psi_{\hat \mu} \gamma_ \mu \psi_{\hat \nu} )$ is the contorsion tensor. The supersymmetry transformations rules are given by
\begin{equation}
\delta_\epsilon e_\mu^{\hat \mu} = \frac \kappa 2 \bar \epsilon \gamma^{\hat \mu} \psi_\mu, \qquad \delta_\epsilon \psi_\mu = \frac 1 \kappa \mathcal D_\mu(\omega)\epsilon
\end{equation}
and $\delta_\epsilon \omega_{\mu {\hat \mu} \hat\nu }$ follows by the chain rule. The asymmetric stress-tensor is given by
\begin{eqnarray}
\theta^{\mu}_{\;\,\nu} &\equiv & e_\nu^{\hat \mu} \frac{\delta \mathcal L_{RS}}{\delta e^{\hat \mu}_\mu} = -\frac i 2 \varepsilon^{\alpha\mu\rho\sigma}\bar \psi_\alpha \gamma_5 (\gamma_\nu D_\rho(\omega) \psi_\sigma+ \Gamma_{\rho\nu}\psi_\sigma)
\end{eqnarray}
and the equations of motion are
\begin{eqnarray}
\label{EOM1}
R^\mu &\equiv& i \varepsilon^{\mu\nu\rho\sigma}\gamma_5 \gamma_\nu \mathcal D_\rho(\omega)\psi_\sigma = 0,\\
\theta_{\mu\nu} &=& \frac e {\kappa^2} (G_{\mu\nu}(\omega)+\Lambda g_{\mu\nu}).\label{EOM2}
\end{eqnarray}

\subsection{Asymptotic solutions}
\label{sec:asol}

We now discuss asymptotic solutions to the equations of motion \eqref{EOM1}, \eqref{EOM2}. We consider general locally asymptotically AdS spacetimes. In standard Fefferman-Graham coordinates, these spacetimes admit metrics near $r\rightarrow \infty$ of the form
\begin{equation}
ds^2 = dr^2 + \gamma_{ij}(x,r)dx^i dx^j, \qquad \gamma_{ij}(x,r) = O(e^{2r}).\label{asympt:metric}
\end{equation}
Accordingly, is natural to work in the ``Fefferman-Graham'' gauge,
\begin{eqnarray}
e^{\hat i}_r &=& 0, \qquad e^{\hat r}_r = 1, \qquad e^{\hat r}_i = 0,\qquad \psi_r = 0 \label{gauge:FG}
\end{eqnarray}
where it turns out that the expansion of asymptotic fields is more easily expressed. In fact, the Fefferman-Graham gauge conditions \eqref{asympt:metric}-\eqref{gauge:FG} need not be enforced to all orders in $r$ for the subsequent asymptotic analysis to be valid. It is sufficient to impose this gauge choice only near the boundary, which still allows ``bulk'' gauge transformations, i.e., transformations not affecting the asymptotic gauge. In an asymptotic expansion near the boundary, four dimensional spinors $\psi_\mu$ can be decomposed into two radially projected spinors $\psi_{\mu,\pm} = P_\pm \psi_\mu$ (which are equivalent to 2-component Weyl spinors in three dimensions), where the projectors are given by $P_{\pm} = \frac 1 2 (1\pm \gamma^{\hat r})$. Note the useful relation $P_\pm \gamma^{\hat i} = \gamma^{\hat i}P_\mp$. The radial coordinate defined by  \eqref{asympt:metric} relates to the coordinates of section \ref{RS} by $r = -\log \Omega$, and in particular the conformal boundary ($\Omega = 0$) now corresponds to $r\to \infty$.

The leading asymptotic behavior $\psi_{i,+}=O(e^{r/2})$ and $\psi_{i,-} = O(e^{-r/2})$ can be inferred from \eqref{asympt:metric} combined with the equations of motion.
The full asymptotic form of the solutions can be obtained by expanding the fields $e^{\hat i}_{i}$, $\psi_{i,\pm}$ in powers of $e^r$ and solving the equations of motion \eqref{EOM1},\eqref{EOM2} order by order in $r$.  There are no power law terms because the theory \eqref{sugraaction} contains a spin-3/2 field with integer-valued mass and the spacetime has an odd-dimensional boundary.  The computation is straightforward but tedious, so we simply state the results below and relegate intermediate asymptotic expansions to appendix~\ref{app:asymptsol}.

The general asymptotic solution is
\begin{eqnarray}
e^{\hat i}_i &=& e^r e_{(0)i}^{\hat i}+ e^{-r} e_{(2)i}^{\hat i} + e^{-2r} e_{(3)i}^{\hat i}+O(e^{-3r}),\nonumber\\
\psi_{i,+} &=& e^{r/2} \psi_{(0)i,+} + e^{-3r/2} \psi_{(3)i,+} +e^{-5r/2} \psi_{(4)i,+} +O(e^{-7r/2}),\label{assol}\\
\psi_{i,-} &=& e^{-r/2} \psi_{(2)i,-} + e^{-3r/2} \psi_{(3)i,-} +O(e^{-5r/2}).\nonumber
\end{eqnarray}
Note that the series for fermions involves only half-integer powers of $e^r$, while the bosonic field has an expansion in integer powers of $e^r$.  This is consistent with the fact that under supersymmetry, a bosonic field transforms into fermion bilinear terms.

We will write the solution in the gauge $e_{(0)}^{i [\hat j} e^{\hat i]}_{(2)i} = 0 = e_{(0)}^{i [\hat j}e^{\hat i]}_{(3)i}$, which can be reached by performing the appropriate local Lorentz transformations (see section \ref{asymptgauge} for details). As in the pure gravity case, the equations of motion imply that the subscript $\mbox{}_{(2)}$ fields are determined in terms of the leading subscript $\mbox{}_{(0)}$ fields:
\begin{eqnarray}
e_{(2)i\hat i} &=& -\frac 1 2 R_{(0)i\hat i}-\frac{i \kappa^2}{4 e_{(0)}}\varepsilon^{kjl}e_{(0)j\hat i} \bar \psi_{(0)k,+}\gamma_5 \gamma_{(0)i}\psi_{(2)l,-}\nonumber\\
&&+e_{(0)i\hat i} \left(\frac{R_{(0)}}{8} + \frac{\kappa^2}{8}\bar \psi_{(0)j,+}\Gamma_{(0)}^{jk}\psi_{(2)k,-}\right),\\
\psi_{(2)i,-} &=& \frac 1 2 \gamma^j_{(0)}D_{(0)j}\psi_{(0)i,+}- \frac 1 2 \gamma^j_{(0)}D_{(0)i}\psi_{(0)j,+} + \frac 1 2 \Gamma_{i(0)}^{\;\, jk}D_{(0)j}\psi_{(0)k,+}.
\end{eqnarray}
The equations of motion also show that the fields $e_{(3)i}^{\hat i}$ and $\psi_{(3)i,-}$ are partially constrained as
\begin{eqnarray}
\gamma^i \psi_{(3)i,-}&=&0,\nonumber \\
0 &=&\varepsilon^{ijk}\gamma_{(0)i}\left( D_{(0)j}\psi_{(3)k,-}- (\frac 3 2 e_{(3)j}^{\hat i}+\frac{\kappa^2}{8}(\bar \psi_{(3)j,-}\psi_{(0)}^{\hat i}-\bar \psi_{(3),-}^{\hat i}\psi_{(0)j}))\gamma_{\hat i}\psi_{(0)k} \right),\nonumber\\
e_{(3)}&=& -\frac{\kappa^2}{6}\bar \psi_{(0)i}\psi_{(3),-}^i,\label{EOMbnd}\\
D^i e_{(3)i}^{\hat i}&=& O(\psi^2),\nonumber
\end{eqnarray}
but still contain degrees of freedom independent of $e_{(0)i}^{\hat i}$ and $\psi_{(0)i,+}$. The $O(\psi^2)$ terms in the above expression are difficult to compute directly from the equations of motion. However, we will derive this divergence of $e_{(3)i}^{\hat i}$  exactly in section~\ref{symmboundary} using another method (namely by requiring boundary diffeomorphism invariance of the action).
All remaining terms in the asymptotic expansions can be determined in terms of $e_{(0)i}^{\hat i}$, $\psi_{(0)i,+}$, $e_{(3)i}^{\hat i}$, and $\psi_{(3),-}$.

We can check that \eqref{assol} is consistent with the linear solution found in section \ref{RS}, where we recall that we imposed the gamma-traceless gauge $\gamma^\mu \psi_\mu = 0$. Beginning with the linear solution \eqref{exp} for $m=1$, we make the appropriate change of radial coordinate and then perform a gauge transformation to set $\psi_r = 0$.  Using the residual gauge freedom, we then send $\psi_i \rightarrow \psi_i + \cD_i \chi$ with $\chi = e^{-r/2}(-\frac 1 3 \gamma^j \psi_{(0)j})+\dots$, so that the linear solution in the Fefferman-Graham gauge \eqref{gauge:FG} is
\begin{eqnarray}
\delta \psi_{i,+ } & = & e^{r/2} \delta \psi_{(0)i,+ } + e^{-3r/2} \delta\psi_{(3)i,+}[\psi_{(0)i,+ }]+ e^{-5r/2} \frac 1 3 \slash \hspace{-0.6em} \partial \delta \psi_{(3)i,-} +O( e^{-7r/2})\\
\delta \psi_{i,-} & = &  e^{-r/2} \delta\psi_{(2)i,-}[\psi_{(0)i,+ }] + e^{-3r/2} \delta \psi_{(3)i,-}+O( e^{-5r/2})\label{linEOM}
\end{eqnarray}
where $\delta\psi_{(2)i,-} = \slash \hspace{-0.6em} \partial \delta \psi_{(0)i,+} - \partial_{i} \delta \cancel \psi_{(0),+} - \frac 1 2 \gamma_i (\partial^k \delta \psi_{(0)k,+} - \slash \hspace{-0.6em} \partial \delta \cancel  \psi_{(0),+})$ and $\delta\psi_{(3)i,+}$ is divergence-free and $\gamma$-traceless. The slash notation is the usual contraction with $\gamma^i$.  The independent fields are indeed $\delta \psi_{(0)i,+}$ and the $\gamma$-traceless and divergence-free $\delta \psi_{(3)i,-}$. For completeness, the linear spin 2 field is given by
\begin{eqnarray}
\delta e^{\hat i}_{i} = \delta e^{\hat i}_{(0)i} e^{r} + \frac 1 2 \d^k \d_k  \delta e^{\hat i}_{(0)i} e^{-r} + \delta e^{\hat i}_{(3)i} e^{-2r} + O(e^{-3r}).
\end{eqnarray}

\subsection{Asymptotic gauge transformations}
\label{asymptgauge}

Having defined the boundary fields in the previous section, we now determine how these fields transform under supersymmetry and more generally under large gauge transformations. In order to do so, we will first have to clarify which gauge transformations preserve the asymptotic form of the metric.

The minimal supergravity Lagrangian is locally gauge invariant under any supersymmetry, diffeomorphism, and local Lorentz transformation triplet $(\epsilon,\xi^\mu,\Lambda^{\hat \nu}_{\;\;\hat \mu})$ acting on the fields as
\begin{eqnarray}
\delta_{\epsilon,\xi,\Lambda} e^{\hat \mu}_\mu &= &\frac{\kappa}{2} \bar \epsilon \, \gamma^{\hat \mu}\psi_{\mu}+ \xi^\alpha \d_\alpha e^{\hat \mu}_{\mu} + \d_\mu \xi^\alpha e^{\hat \mu}_\alpha + \Lambda^{\hat \mu}_{\;\;\hat \nu} e^{\hat \nu}_\mu,\label{gaugegeneral}\\
\delta_{\epsilon,\xi,\Lambda}\psi_\mu &=& \frac{1}{\kappa} (D_\mu \eps - \frac 1 2 \gamma_\mu \eps) + \xi^\alpha \d_\alpha \psi_\mu + \d_\mu \xi^\alpha \psi_\alpha + \frac 1 4 \Lambda_{\hat \mu \hat \nu}\Gamma^{\hat \mu \hat \nu}\psi_\mu \,.
\end{eqnarray}
Let us now determine the subset of these transformations that preserves the asymptotic conditions \eqref{gauge:FG} near $r \rightarrow \infty$.

The condition $\delta_{\epsilon,\xi,\Lambda} e^{\hat r}_r = 0$ is satisfied only if $\xi^r = \xi_{(0)}^r(x^i)$. The condition $\delta_{\epsilon,\xi,\Lambda} e^{\hat r}_i = 0$ then implies that a compensating local Lorentz transformation
\begin{equation}
\Lambda^{\hat r}_{\;\;\hat i} = - \frac \kappa 2 \bar \eps \gamma^{\hat r}\psi_j e^j_{\hat i} - \d_i \xi_{(0)}^r e^i_{\hat i} \label{compLor}
\end{equation}
has to be performed for each supersymmetry transformation and radial diffeomorphism. The condition $\delta_{\epsilon,\xi,\Lambda} e^{\hat i}_r = 0$ determines the behavior of $\xi^i$ up to an arbitrary integration function,
\begin{equation}
\xi^i = \xi^i_{(0)}(x^j) - \frac \kappa 2 \int dr \, \bar \epsilon \gamma^{\hat r} \gamma^{ij} \psi_j -  \int dr \, \gamma^{ij}\d_j \xi^r_{(0)}\,.\label{comp:diff}
\end{equation}
The last condition $\delta_{\epsilon,\xi,\Lambda} \psi_r = 0$ is particularly interesting since it requires the supersymmetry parameter to be a function of only two projected spinors $\eps_-(x^i)$ and $\eps_+(x^i)$.  A detailed analysis (using the results in appendix \ref{app:asymptsol}) leads to the expansion
\begin{eqnarray}
\epsilon &=&  \eps_+ e^{r/2} + \eps_- e^{-r/2} + e^{-3r/2}\eps_{(3),+}[\eps_-,\eps_+,\xi^r] +\nonumber\\
&& e^{-5r/2} (\eps_{(4),+}[\eps_+] + \eps_{(4),-}[\eps_-,\eps_+,\xi^r]) + O(e^{-7r/2})\,.\label{susyexp}
\end{eqnarray}
The full expressions for $\eps_{(3),+},\eps_{(4),+}, \eps_{(4),-}$ are given in appendix \ref{app:asymptsol}.
Realizing that $\eps_{(3),-} = 0$ and that the term $\eps_{(4),+}[\eps_+]$ has a simple expression in terms of only $\eps_+$ will be crucial to describing how $\delta_{\epsilon,\xi,\Lambda}$ act on the fields $e_{(3)i}^{\hat i}$ and $\psi_{(3),-}$. Local Lorentz transformations preserving the asymptotic behavior of the fields $e^{\hat i}_i$, $\psi_{i,\pm}$ have the form $ \Lambda^{\hat i}_{\;\; \hat j}(r,x^i)= O(e^{0r})$.

At this step, we can summarize the only transformations $(\eps,\xi,\Lambda)$ preserving the asymptotic structure (in addition to transformations deep in the bulk, i.e. outside the asymptotic region). We found only the diffeomorphisms generated by $\xi^i_{(0)}$, the radial diffeomorphisms generated by $\xi^r_{(0)}$, the two supersymmetries $\eps_-(x^i)$ and $\eps_+(x^i)$, and any local Lorentz transformation $\Lambda^{\hat i}_{\;\; \hat j}(r,x^i)= O(e^{0r})$.

Let us now derive the action of these generators on the boundary fields. It is obvious from \eqref{gaugegeneral} that $\xi = \xi^i_{(0)}(x^j)$ generates diffeomorphisms at the boundary. The radial diffeomorphisms generated by $\xi^r = \xi^r_{(0)}(x^j)$ act as simple Weyl transformations on the boundary fields with the expected conformal weights,
\begin{eqnarray}
\delta_{\xi^r_{(0)}}  e^{\hat i}_{(0)i} &=& e^{\hat i}_{(0)i} , \qquad
\delta_{\xi^r_{(0)}}  \psi_{(0)i,+} = \frac 1 2 \psi_{(0)i,+} \\
\delta_{\xi^r_{(0)}}   e^{\hat i}_{(3)i}&= & - 2 e^{\hat i}_{(3)i},\qquad
\delta_{\xi^r_{(0)}}  \psi_{(3)i,-} = -\frac 3 2 \psi_{(3)i,-} \, .
\end{eqnarray}
Note that the gauge transformations induced by the generators \eqref{compLor}, \eqref{comp:diff}, \eqref{susyexp} (which  depend on $\xi^r$) do not contribute to the above result. Let us expand the local Lorentz transformations as
\begin{eqnarray}
\Lambda^{\hat i}_{\;\; \hat j}(r,x^i)= \Lambda^{\hat i}_{(0)\hat j}+\Lambda^{\hat i}_{(2)\hat j} e^{-2r} + \Lambda^{\hat i}_{(3)\hat j} e^{-3r}+ O(e^{-4r}).
\end{eqnarray}
Here we have neglected the term of order $e^{-r}$, which is not needed in what follows.  The transformations induced by $\Lambda^{\hat i}_{(0)\hat j}$,
\begin{eqnarray}
\delta_{\Lambda_{(0)}}  e^{\hat i}_{(0)i} &=& \Lambda^{\hat i}_{(0)\hat j} e^{\hat j}_{(0)i} , \qquad
\delta_{\Lambda_{(0)}}  \psi_{(0)i,+} = \frac 1 4 \Lambda_{(0)\hat i\hat j} \Gamma^{\hat i\hat j}\psi_{(0)i,+} \\
\delta_{\Lambda_{(0)}}   e^{\hat i}_{(3)i}&= & \Lambda^{\hat i}_{(0)\hat j} e^{\hat j}_{(3)i},\qquad
\delta_{\Lambda_{(0)}}  \psi_{(3)i,-} = \frac 1 4 \Lambda_{(0)\hat i\hat j} \Gamma^{\hat i\hat j}\psi_{(3)i,-} \, ,
\end{eqnarray}
show that all boundary fields transform as vectors or spinors under the Lorentz transformation $\Lambda^{\hat i}_{(0)\hat j}$. However, the transformation $\delta_{\Lambda_{(2)},\Lambda_{(3)}}$ acts non-trivially on the fields
\begin{eqnarray}
\delta_{\Lambda_{(2)},\Lambda_{(3)}}  e^{\hat i}_{(2)i} &=& \Lambda^{\hat i}_{(2)\hat j} e^{\hat j}_{(0)i} , \qquad
\delta_{\Lambda_{(2)},\Lambda_{(3)}}   e^{\hat i}_{(3)i}=  \Lambda^{\hat i}_{(3)\hat j} e^{\hat j}_{(0)i}.\label{gauge3}
\end{eqnarray}
while $e^{\hat i}_{(0)i}$, $\psi_{(0)i,+}$, $\psi_{(2)i,-}$, and $\psi_{(3)i,-}$ are invariant. These gauge transformations can be used to fix the gauge $e_{(0)}^{i [\hat j} e^{\hat i]}_{(2)i} = 0 = e_{(0)}^{i [\hat j}e^{\hat i]}_{(3)i}$, as done in section \ref{sec:asol}.

Finally, we address the supersymmetry transformations. A careful expansion of \eqref{gaugegeneral} taking into account the indirect contributions of the other gauge generators \eqref{compLor}, \eqref{comp:diff}, \eqref{susyexp} leads to the result
\begin{eqnarray}
\delta_{\eps_\pm} e^{\hat i}_{(0)i} &=& \frac{\kappa}{2} \bar \eps_+ \gamma^{\hat i}\psi_{(0)i,+},\nonumber \\
\delta_{\eps_\pm} \psi_{(0)i,+} &=& \frac{1}{\kappa} D^{(0)}_{i}\eps_{+} - \frac{1}{\kappa} \gamma_{\hat i} e^{\hat i}_{(0)i} \eps_-,\nonumber \\
\delta_{\eps_\pm}  e^{\hat i}_{(3)i}&= & \frac{\kappa}{2}\bar \eps_- \gamma^{\hat i} \psi_{(3)i,-} - \frac{\kappa}{6} \cL_{\bar \eps_+ \psi_{(3),-}} e_{(0)i}^{\hat i} + \frac{\kappa}{2}\bar \eps_+ \gamma^{\hat i}\psi_{(4)i,+} + \frac{\kappa}{2} \bar \eps_{(4),+}\gamma^{\hat i}\psi_{(0)i,+}, \label{susynonlin}\\
\delta_{\eps_\pm} \psi_{(3)i,-} &=& -\frac{3}{2\kappa}\gamma_{\hat i} e^{\hat i}_{(3)i}\eps_{+}- \frac{\kappa}{8}(\psi_{(3)i,-}\psi^{\hat j}_{(0),+}- \psi_{(0)i,+}\psi^{\hat j}_{(3),-})\gamma_{\hat j}\eps_+ - \frac{\kappa}{4}\bar \eps_+ \psi_{(3),-}^{\hat j} \gamma_{\hat j}\psi_{(0)i,+}\, ,\nonumber
\end{eqnarray}
where $\psi_{(4)i,+} $ and $\eps_{(4),+}$ are given in \eqref{psi4}, \eqref{eps4}. One readily identifies that $(e^{\hat i}_{(0)i},\psi_{(0)i,+})$ form a $\cN = 1$ $d=3$ superconformal multiplet \cite{Achucarro:1987vz}. The transformations $\eps_+$ generate two (real) supersymmetries, while $\eps_-$ generate the two (real) special conformal supersymmetries. This result is the  AdS$_4$ counterpart of the results obtained in AdS$_3$ \cite{Nishimura:1998ud}, AdS$_5$ \cite{Liu:1998bu,Balasubramanian:2000pq} and AdS$_6$ \cite{Nishimura:2000wj}.

Using an additional gauge transformation \eqref{gauge3} with the parameter $\Lambda^{\hat i}_{(3)\hat j} = - \frac \kappa 6 \bar \eps_+ \psi_{(3),-}^j \omega_{j\;\;\,\hat j}^{\;\hat i}+\frac {\kappa}{4}\bar \eps_+ \Gamma^{\hat i \hat j}\gamma^l \psi_{(0)l,+}$, the transformation of the fields $(e^{\hat i}_{(3)i},\psi_{(3)i,-})$ at lowest order in the number of fermions is more easily expressed as
\begin{eqnarray}
\delta_{\eps_\pm,\Lambda_{(3)}}  e^{\hat i}_{(3)i}&= & \frac{\kappa}{2}\bar \eps_- \gamma^{\hat i} \psi_{(3)i,-} +\frac \kappa {6} \bar \eps_+ \gamma^{\hat i}\slash \hspace{-0.6em}  D_{(0)}   \psi_{(3)i,-}- \frac{\kappa}{6} D_{(0)i}(\bar \eps_+ \psi_{(3)j} e^{\hat i \,j}_{(0)})\nonumber\\
&& - \frac\kappa 4 \bar \eps_+ \gamma^l \psi_{(0)l}e_{(3)i}^{\hat i}+ \frac \kappa 2 \bar \eps_+ \gamma^{\hat i}\psi_{(0)}^{\hat j}e_{(3)i\hat j}+O(\psi^3),\label{eq:45}\\
\delta_{\eps_\pm} \psi_{(3)i,-} &=& -\frac{3}{2\kappa}\gamma_{\hat i} e^{\hat i}_{(3)i}\eps_{+}+O(\psi^2)\,.\nonumber
\end{eqnarray}
The fields $(e^{\hat i}_{(3)i},\psi_{(3)i,-})$ transform linearly under the supersymmetry transformations.

\subsection{Conjugate fields and regularization of the action}
\label{regaction}

The supergravity action \eqref{sugraaction} is infinite on-shell and does not lead to a well-defined variational principle. Both problems can be cured by a regularization procedure systematically developed under the name of holographic renormalization \cite{deHaro:2000xn,Bianchi:2001kw,Bianchi:2001de} (see also the Hamiltonian \cite{Papadimitriou:2004ap} and Hamilton-Jacobi \cite{deBoer:1999xf} approaches). As detailed for example in \cite{Andrade:2006pg}, appropriate boundary terms must exist such that the total action $S_{reg} = \int_{\cM} d^{\,4}x \mathcal L_{sugra} + \int_\cI d^3x \cL_{bosonic,ct} + \int_\cI  d^3x \cL_{fermionic,ct}$ admits a variation of the form
\begin{equation}
\delta S_{reg} = \int_{\cI} d^3x\,\left( e_{(0)} T^{i\hat i} \delta e_{(0)i\hat i}+ e_{(0)} \bar J^i \delta \psi_{(0)i}\right) \,,\label{var:final}
\end{equation}
where $T^{i\hat i}$ is the finite stress-tensor and $J^i$ is the finite supercurrent (conjugates to  the boundary vielbein and Rarita-Schwinger field, respectively). For Einstein gravity, the counterterms $\int_\cI \cL_{bosonic,ct}$ are given by \cite{Balasubramanian:1999re}
\begin{equation}
 \int_\cI \cL_{bosonic,ct} = \frac{1}{\kappa^2} \int_\cI e K - \frac{2}{\kappa^2} \int_\cI   e + \frac{1}{2\kappa^2}\int_\cI  e R . \label{bosonicc}
\end{equation}
We can attempt to infer the fermionic contributions to the boundary term from the linear results obtained in section \ref{renorm}. Coupling the counterterm \eqref{fermcounter} for $d=4$, $m=1$, $N=-1/2$ to the vielbein $e$ and the connection $\omega$, we obtain
\begin{equation}
 \int_\cI \cL_{fermionic,ct} = -\frac 1 4 \int_\cI  e \left( \bar \psi_i \psi^i -2 \bar \psi_i \gamma^j D_j \psi^i \right) \,, \label{fermionicc}
\end{equation}
where $D$ is the covariant derivative associated with $\gamma_{ij}$.  These counterterms contribute a finite piece
\begin{equation}
 \int_\cI \cL_{ct,finite} = -\frac{2}{\kappa^2} \int_\cI e_{(0)}e_{(3)}-\frac{1}{2} \int_\cI e_{(0)}\bar \psi^i_{(0),+}\psi_{(3)i,-} \,\label{eq:finitepartcount}
\end{equation}
such that the variation of the regularized action is indeed given by an expression of the form \eqref{var:final}.  It is reasonable to expect that the counterterms \eqref{bosonicc} and \eqref{fermionicc} are sufficient to cancel all divergences appearing in the expansion of $\delta \int_{\cM} d^4x \mathcal L_{sugra} = \int d^{\,3}x \Theta^r$ with $\Theta^r = \frac{1}{\kappa^2}e e^{i \hat i}\delta \omega_{i \hat r \hat i} + \frac{1}{2}\bar \psi_i \Gamma^{ij}\gamma^{\hat r}\delta \psi_j$.  However, this has not been explicitly verified due to the large number of terms generated.

With the above choice of counterterms, the stress-tensor and the supercurrent are found to be
\begin{eqnarray}
T^{i\hat i} &=& \frac{3}{\kappa^2}e_{(3)}^{\hat i i} + \bar \psi^{(i}_{(0)}\psi_{(3)}^{\hat i)}-\frac 1 2 \bar \psi_{(0)}^k \gamma_k \gamma^{i}\psi_{(3)}^{\hat i},\label{def:Tij}\\
J^i &=& - \psi_{(3)}^{i}\, .
\end{eqnarray}
These fields obey the following constraints obtained from the bulk equations of motion (see \eqref{EOMbnd}):
\begin{eqnarray}\label{conditions:stress}
\gamma^i J_i &=& 0,\qquad e^{i \hat i}_{(0)} T_{i\hat i} + \frac 1 2 \bar J^i \psi_{(0)i}= 0,\label{constr1}\\
D^{(0)}_i(\omega) J^i &=& \frac{\kappa^2}{2}T^{ij} \gamma_j \psi_{(0)i,+}, \qquad D_{(0)}^i T_{i\hat i}= O(\psi^2).\nonumber
\end{eqnarray}

\subsection{Boundary gauge invariance}
\label{symmboundary}

The supergravity action \eqref{sugraaction} is gauge invariant under diffeomorphisms and supersymmetry transformations only up to boundary terms. We have shown that a generic variation of the regularized action can be written on-shell as \eqref{var:final}. Using the constraints \eqref{conditions:stress}, one can now analyze if the action is invariant under all gauge transformations \emph{also at the conformal boundary}\footnote{We do not consider the boundary at initial and final times in this paper.}.

Varying the regularized action under the supersymmetry transformations \eqref{susynonlin}, we obtain
\begin{equation}
\delta S_{reg} = \int_{\cI} d^{\,3}x\,\left[e_{(0)} \bar\eps_+ \left(   \frac{\kappa}{2}T^{i\hat i} \gamma_{\hat i} \psi_{(0)i,+} - \frac 1 \kappa D^{(0)}_i(\omega)J^i \right) + \frac 1 \kappa e_{(0)} \bar\eps_-  \gamma_i J^i \right]+ \frac 1 \kappa \int_{\partial\cI} d^{\,2} x e_{(0)} \bar \eps_+ J^i N_i  \,,\label{var:susy}
\end{equation}
where $N^i$ is the normal at the boundary $\partial\cI$ of the conformal boundary $\cI$. Thus, we see that the constraints \eqref{constr1} imply the invariance of the regularized action under supersymmetries and special conformal symmetries, up to a possible term at the boundary of $\cI$. The boundary of a constant time slice of AdS in global coordinates is just a sphere without boundary, but one may consider, for example, AdS in a Poincar\'e patch whose flat boundary has a spatial as well as null infinity.  If $\partial\cI$ is not empty, we conclude that further care has to be taken to study these boundaries; we will not pursue such an analysis here. To summarize, we have shown that the action is invariant under supersymmetry, up to boundary terms of codimension 2.

The supersymmetry transformations of the stress-tensor and the supercurrent are straightforward consequences of \eqref{eq:45} and the definition \eqref{constr1}. A direct calculation yields
\begin{eqnarray}\label{susyTJ}
\delta_{\eps_\pm,\Lambda_{(3)}} T^{i\hat i} &=& \frac 1 \kappa (2 J^i \gamma^{\hat i}-J^{\hat i}\gamma^i)\eps_- - \frac{1}{2\kappa}\bar \eps_+ \Gamma^{\hat i j}D^{(0)}_j J^i + \frac{1}{2\kappa}\bar J^{\hat i}\Gamma^{ij}D^{(0)}_j \eps_+ \nonumber\\
&& (T \cdot\psi_{(0)} \cdot \eps_+)\text{-terms} + (J \cdot\psi_{(0)} \cdot\psi_{(0)}\cdot \eps_+)\text{-terms}\label{susy:trans} \\
\delta_{\eps_\pm} J^i &=& \frac{\kappa}{2}T^{i\hat i}\gamma_{\hat i}\eps_+ + (J \cdot \psi_{(0)} \cdot \eps_+)\text{-terms}\nonumber
\end{eqnarray}
We have computed explicitly the terms of higher order in fermions, but we will only need in what follows that such terms are linear in the stress-tensor or the supercurrent. Another route to deriving these transformations consists of acting on both sides of \eqref{var:final} with a supersymmetry transformation $\delta_{\eps_\pm}$. Upon commuting $\delta$ with
$\delta_{\eps_\pm}$, we note that the left-hand side vanishes, while the right-hand side takes the form $(\delta_{\eps_\pm} T^{i\hat i} + \dots )\delta e_{(0)i\hat i} + (\delta_{\eps_\pm} \bar J^i +\dots )\delta \psi_{(0)i,+}$.  Setting the coefficients of $\delta e_{(0)i\hat i}, \delta \psi_{(0)i,+}$ to zero then reproduces the above results.

One can similarly prove the exact invariance of the regulated action under boundary conformal transformations using \eqref{constr1}. Proving the invariance under boundary diffeomorphisms directly would require the $O(\psi^2)$ terms that we were unable to find in closed form in \eqref{EOMbnd}. However, the successive action of two supersymmetries is just a combination of a diffeomorphism and other gauge transformations. Our previous checks of invariance thus also prove boundary diffeomorphism invariance. Therefore, we can use diffeomorphism invariance to compute the missing term in \eqref{EOMbnd}.

We consider a combination of a boundary diffeomorphism together with a local Lorentz transformation chosen to maintain the gauge
\begin{equation}
\delta_\xi e_{(0)i \hat i} = \frac{1}{2}  e_{(0)}^j{}_{\hat i} \delta_\xi g_{(0) i j} = e_{(0)}^k{}_{\hat i} D^{(0)}_{(i} \xi_{k)} \,.
\end{equation}
Plugging this variation in \eqref{var:final} and requiring that the boundary term vanishes for all diffeomorphisms $\xi_{(0)}^i$ then requires
\begin{eqnarray}\label{conditions:stress2}
D_{(0)}^i T_{(i k)} = \overline{J^i} D^{(0)}_k \psi_{(0) i }-D_{(0)}^i \left(\overline{J_i}  \psi_{(0) k}\right)
+\frac{1}{4} D_{(0)}^l \left(\overline{J^i}  \Gamma^{(0)}_{k l } \psi_{(0) i}\right)
- \frac{1}{4} \kappa^{(0)}_{k}{}^{ j l} \overline{J^i}  \Gamma^{(0)}_{j l } \psi_{(0) i} \,.
\end{eqnarray}
Using the definition of the stress-tensor \eqref{def:Tij}, one can finally obtain the missing piece in \eqref{EOMbnd}.

\subsection{Boundary conditions}
In this section we determine valid boundary conditions that are consistent with supersymmetry.
The action $S_{Dir} \equiv S_{reg}$ provides a variational principle \eqref{var:final} suitable to impose Dirichlet boundary conditions on the boundary fields $(e_{(0)i}^{\hat i},\psi_{(0)i,+})$.  It is clear from the transformations \eqref{susynonlin} that these boundary conditions are preserved under supersymmetry for supersymmetry parameters obeying the background Killing spinor equations.  This variational principle is interpreted in AdS/CFT language as specifying the values of the sources dual to the boundary operators.

The linear analysis performed in section \ref{RS} reached the conclusion that Neumann boundary conditions exist for the massless Rarita-Schwinger fields appearing in supergravity. Now, it is straightforward to extend this analysis to the non-linear case by considering the Legendre-transformed action
\begin{equation}
S_{Neu} = S_{Dir} - \int_\cI d^3 x \left( e_{(0)}T^{i \hat i } e_{(0)i\hat i} + e_{(0)}\bar J^i \psi_{(0)i,+}\right) \,,
\end{equation}
which is stationary under the Neumann boundary conditions $\delta(e_{(0)}T^{i \hat i }) = \delta (e_{(0)}\bar J^i) = 0$. It follows from the linearity of the supersymmetry transformations \eqref{susy:trans} in the stress-tensor and the supercurrent that the particular boundary condition
\begin{equation}
T^{i \hat i } = 0, \qquad J^i = 0\, \label{NeuBC}
\end{equation}
is preserved under supersymmetry and special conformal supersymmetry transformations. Because the Neumann action reduces to the Dirichlet action upon imposing the boundary conditions \eqref{NeuBC}, it follows from section \ref{symmboundary} that $S_{Neu}$ (with \eqref{NeuBC}) is also gauge invariant under all local diffeomorphism, Weyl, and supersymmetry transformations.  It is clear from \eqref{flux} that these Neumann boundary conditions conserve the renormalized symplectic structure.  Based on the corresponding results for pure gravity \cite{Compere:2008us} and linearized spin-3/2 fields (Section \ref{renorm}), we conjecture that the renormalized symplectic structure for the Neumann theory is indeed finite, though we leave rigorous verification of this for future work. Following \cite{Compere:2008us} (see also \cite{Hollands:2006zu}), it should also be possible to show that the charges associated with the gauge symmetries vanish for the boundary conditions \eqref{NeuBC}.

The theory described by four-dimensional supergravity with boundary conditions \eqref{NeuBC} can therefore be thought of as a three-dimensional conformal supergravity theory.  In the context of the AdS/CFT correspondence, this theory has been interpreted as the induced gravity theory of the dual CFT \cite{Compere:2008us}.

Relevant or marginal deformations of the boundary CFT correspond to adding a boundary action to $S_{Neu}$ as
\begin{equation}
S_{Neu,def} = S_{Neu} + S_{bnd}[e_{(0)i}^{\hat i},\psi_{(0)i,+}] - \varQ{S_{bnd}}{e_{(0)i}^{\hat i}}e_{(0)i}^{\hat i} - \varQ{S_{bnd}}{\psi_{(0)i,+}}\psi_{(0)i,+},\label{totaction}
\end{equation}
where $S_{bnd}[e_{(0)i}^{\hat i},\psi_{(0)i,+}] = \lambda_{(0)} \int_\cI d^3x \mathcal L_{bnd}$ contains only terms with mass dimension less than or equal to $3$. The general boundary action satisfying such requirements and admitting $\cN = 1$ supersymmetry is the supersymmetric cosmological topologically massive gravity (SCTMG) \cite{Deser:1982sw,Deser:1982sv} (see also \cite{Carlip:2008eq,Gibbons:2008vi}),
\begin{eqnarray}
\mathcal L_{bnd} &=& \frac{1}{2 \kappa_{(0)}^2} e_{(0)} \,(R_{(0)}-2\Lambda_{(0)}) - \frac{e_{(0)} }{2} \left( \bar \psi_{(0)i} \Gamma_{(0)}^{ijk} D^{(0)}_j \psi_{(0)k} +\frac{i}{2 l_{(0)}}\bar \psi_{(0)i}\gamma_5\Gamma_{(0)}^{ij}\psi_{(0)j}\right)+\mathcal L_{top}\nonumber \\
&=& \frac{1}{2 \kappa_{(0)}^2} e_{(0)} \,(R_{(0)}-2\Lambda_{(0)}) + \frac{i }{2} \varepsilon^{ijk} \bar \psi_{(0)i} \gamma_5 \left( D^{(0)}_j \psi_{(0)k} + \frac{i}{2 l_{(0)}} \gamma_5\gamma_{(0)j} \psi_{(0)k}\right)+\mathcal L_{top},
\end{eqnarray}
where $\Lambda_{(0)}=-1/l^2_{(0)}$, $\varepsilon^{012} = +1$, and $\epsilon^{rijk}=\epsilon^{ijk}$ ($\hat r=2$). The unusual $\gamma_5$ matrices appear because we are using a four-dimensional representation of the three-dimensional Clifford algebra. The ``topological'' term is
\begin{eqnarray}
\mathcal L_{top} &=& \frac{1}{8\kappa_{(0)}^2\mu}\int \varepsilon^{ijk}\left( R_{ij}^{(0)\;\hat i \hat j}\omega_{(0)k\hat i\hat j}+\frac 2 3 \omega^{(0)\hat i \hat j}_{i}\omega_{j \hat i}^{(0)\;\, \hat k}\omega_{(0)k\hat k \hat j} \right)+\frac{i\mu}{4} \int e^{-1}_{(0)}\bar R_{(0)}^i \gamma_5\gamma_{(0)j} \gamma_{(0)i} R_{(0)}^j \nonumber
\end{eqnarray}
where $R_{(0)}^i = \varepsilon^{ijk}D_{(0)j}\psi_{(0)k}$.

It is important to note that the spin connection $\omega_{(0)i}^{\hat i \hat j}$ is not an independent field in this action. By construction of the action, it is a solution of $D^{(0)}_{[\mu}(\omega_{(0)}) e^{(0)\hat \mu}_{\nu]} = \frac{\kappa_{(0)}^2}{4}\bar \psi^{(0)}_\mu \gamma^{\hat \mu} \psi^{(0)}_\nu $. Now, the latter equation is precisely the leading order behavior of the definition of the spin connection in the bulk \eqref{eq:spin}. To have a consistent coupling of the bulk and boundary theories, we are forced to impose $\kappa_{(0)} = \kappa$. It will be seen below that this is also required for supersymmetry. The total action \eqref{totaction} then provides a valid variational principle for the boundary conditions
\begin{eqnarray}
T^{i}_{\;\hat i} + \frac{1}{e_{(0)}} \varQ{\cL_{bnd}}{e_{(0)i}^{\hat i}} = 0, \qquad \bar J^i+\frac{1}{e_{(0)}} \varQ{\cL_{bnd}}{\psi_{(0)i,+}}=0,\label{eq:mix}
\end{eqnarray}
which are namely the equations of motion of SCTMG coupled to the boundary stress-tensor and supercurrent.

Let us now study the gauge symmetries of the deformed theory. First, the Einstein-Hilbert and cosmological term break Weyl invariance. The topologically massive gravity is invariant under the supersymmetry transformations
\begin{eqnarray}
\delta_{\alpha} e^{\hat i}_{(0)i} &=& \frac{\kappa_{(0)}}{2} \bar \alpha  \gamma^{\hat i}\psi_{(0)i},\\
\delta_{\alpha} \psi_{(0)i} &=& \frac{1}{\kappa_{(0)}}(D^{(0)}_{i}  + \frac{i}{2 l_{(0)}} \gamma_5\gamma_{(0)i}) \alpha .
\end{eqnarray}
In order to obtain a supersymmetric deformation, the crucial point is to match these supersymmetry transformations with the transformations following from the bulk action \eqref{susynonlin}. We observe that these conditions are met only for the following particular combination of the supersymmetries
\begin{eqnarray}
\eps_{+} = \alpha, \qquad \eps_- = \frac{i}{2 l_{(0)}}\gamma_5 \alpha \label{alpha}\, .
\end{eqnarray}
It then follows from arguments similar to those given in the pure Neumann case that the mixed boundary conditions \eqref{eq:mix} are also preserved by the supersymmetry transformations \eqref{susynonlin}. Therefore, we conclude that there exists a family  of relevant and marginal deformations of the Neumann theory characterized by the 3-parameters $(\lambda_{(0)},l_{(0)},\mu)$.  The boundary supersymmetry is a particular combination of large bulk supersymmetries that preserve the Fefferman-Graham asymptotic expansion and depend on the value of the boundary cosmological constant $l_{(0)}$.

In general, these deformations preserve $\cN = 1$ supersymmetry but break special conformal supersymmetry and Weyl invariance.  In the special case where there is no boundary cosmological term or boundary Einstein term, the gauge symmetries are enhanced to the full $\cN = 1$ superconformal invariance since the marginal ``topological term'' is itself invariant under Weyl rescaling and under any supersymmetry transformation \eqref{susynonlin}.

\section{Discussion}
\label{disc}

In this work, we have investigated linear Rarita-Schwinger fields in Lorentzian AdS$_d$ with a mass term $m \bar \psi_\mu \Gamma^{\mu\nu}\psi_\nu$. This analysis covered in particular the gauge invariant ``massless'' fields with $m = \frac{d-2}{2 l_{AdS}}$, which are relevant to the massless multiplet in supergravity theories.  Under the standard notion of normalizeability, Rarita-Schwinger fields in AdS admit mixed Dirichlet-Neumann boundary conditions in the mass range $0 \leq |m| < \frac 1{2l_{adS}}$, matching exactly the corresponding range for spin-1/2 fields \cite{Amsel:2008iz}. For the mass range $|m| \geq  \frac{1}{2 l_{AdS}}$, which includes the ``massless'' Rarita-Schwinger fields, only  Dirichlet conditions are allowed according to the standard notion of normalizeability.

 We argued that for any mass, the Rarita-Schwinger action can be ``renormalized,'' so that it is finite and obeys a well-defined variational principle.  According to the procedure developed in \cite{Compere:2008us},  it should therefore be possible to renormalize accordingly the symplectic structure and allow Neumann-type boundary conditions for any mass.  The minimal counterterms have been computed up to $|m| \leq \frac{3}{2 l_{AdS}}$, and normalizeability was explicitly checked for the parameter range $1/2 < m \leq 1$, $d \geq 3$, which includes the particularly interesting case of massless fields in $ d= 4$.

 The results for massless Rarita-Schwinger fields in $d = 4$ have been used to explore supersymmetric boundary conditions for $\cN = 1$ AdS$_4$ supergravity in which the metric and Rarita-Schwinger fields are fluctuating at the boundary. Using the AdS/CFT dictionary, Neumann boundary conditions in $d=4$ supergravity correspond to gauging the superconformal group of the 3-dimensional CFT describing M2-branes and path-integrating the boundary supergravity multiplet. Supersymmetry and conformal supersymmetry have been shown to be exact gauge invariances of the renormalized action. This implies that the boundary Neumann theory is supersymmetric and conformally invariant in the large $N$ approximation. We notably obtained the expressions for the boundary stress-tensor multiplet and its supersymmetry transformations. We pointed out that relevant and marginal $\cN = 1$ supersymmetric deformations of the induced supergravity theory are obtained by coupling the theory to $\cN = 1$ superconformal topologically massive gravity. In general, Weyl invariance is then broken, but $\cN = 1$ supersymmetry is preserved.

It is important to address whether or not the Neumann theories are unitary. While it is difficult to conclusively answer this question in the affirmative, we can at least check certain necessary conditions for unitarity, namely the unitary bounds obeyed by CFT operators dual to the bulk fields (see e.g. \cite{Klebanov:1999tb,Minwalla:1997ka,Grinstein:2008qk}).  We now address this issue in some detail for theories of linearized bulk fields.

In a $n \equiv d-1$ dimensional CFT, a scalar operator is required to have a conformal dimension bounded from below as $\Delta \geq \frac{n-2}{2}$ so that none of its descendants have negative norm\footnote{There are, however, some cases of scalar operators with $\Delta = 0$ that preserve unitarity \cite{Minwalla:1997ka}.}. Using the AdS/CFT dictionary, scalar fields of mass $m^2$ are dual to operators of dimension $\Delta_\pm (\Delta_\pm - n) = m^2$, depending if one imposes Dirichlet ($\Delta_+$) or Neumann ($\Delta_-$) boundary conditions. Operators of dimension $\Delta_+$ always obey the unitary bound, while operators of dimension $\Delta_-$ are consistent with unitarity only for $m^2 \leq -\frac{n^2}{4}+1$. This upper bound for the mass corresponds in the bulk to the upper limit of the Breitenlohner-Freedman mass range.  More precisely, the condition for both linearly independent modes to be normalizeable is $m^2 < -\frac{n^2}{4}+1$, so it may be interesting to study the borderline case $m^2 =-\frac{n^2}{4}+1$ (when the unitarity bound is saturated) more carefully.

Constraints on spin-1/2 fields in supersymmetric CFTs were obtained in \cite{Minwalla:1997ka}. Spin-1/2 primary fields should have a dimension bounded by $\Delta \geq \frac{n-1}{2}$. Using the AdS/CFT dictionary, we have $\Delta_\pm = \frac{n}{2}\pm |m|$. If the dual CFT operator is a primary spin-1/2 operator or a descendant thereof, we then see that the dimension of the dual CFT operator in the Dirichlet theory always satisfies the unitarity bound, while for the Neumann theory the bound implies $|m| \leq \frac{1}{2}$.   This bound corresponds in the bulk to the normalizeability bound $|m| < \frac{1}{2}$ \cite{Amsel:2008iz} (except once again for the borderline case $|m| = \frac{1}{2}$).

For spins $s \geq 1$, the analysis becomes more involved due to the presence of gauge freedom. \emph{Gauge invariant} primary vector operators obey the unitarity bound $\Delta \geq n -1$. For Maxwell theory in AdS with Dirichlet boundary conditions, this bound is saturated for the conserved current dual to the boundary vector field. In the Neumann theory, the dual operator is a vector potential of dimension $\Delta = 1$, which appears to violate the bound. However, the unitarity bound actually does not apply since the boundary operator is not  gauge invariant.  The analysis of massless spin-3/2 and spin-2 fields is quite similar because once again there is gauge freedom, namely the fermionic supersymmetry-type transformations or diffeomorphisms.  For Neumann boundary conditions, the operator dual to the source in the bulk is not gauge invariant, and therefore, unitarity of these theories cannot be ruled out using the unitarity bounds derived in \cite{Minwalla:1997ka}. For massive spin-3/2 theories, the dimensions of Dirichlet and Neumann operators are given by $\Delta_\pm = \frac{n}{2} \pm |m|$ and the bound for gauge invariant primary operators obtained in supersymmetric CFTs is $\Delta \geq n - \frac{1}{2}$. We can apply the unitarity bounds only if the operators dual to the sources are spin-3/2 primary operators or descendants of spin-3/2 operators, though it is not clear whether or not this hypothesis is justified.  Assuming that this statement does hold for the sake of argument, the unitarity bounds imply that Neumann boundary conditions for massive Rarita-Schwinger fields lead to ghosts (i.e., violation of unitarity).  Moreover, in contrast to the lower spin cases, the unitarity bound implies a \emph{lower} bound on the mass of the Rarita-Schwinger field for the \emph{Dirichlet} theory. In particular, the mass has to be greater than the ``massless'' value, $|m| \geq m_0 = \frac{n-1}{2}$. This is satisfied, for example, for spin-3/2 fields appearing in Type IIB supergravity on $AdS^5 \times S^5$ \cite{Kim:1985ez}.

In summary, the unitarity of the Neumann theory of massless Rarita-Schwinger fields ($m = \frac{d-2}{2}$) cannot be ruled out by simple application of the unitary bounds of the dual CFT. Now, we have seen in section \ref{prop} that Rarita-Schwinger fields with masses $m = (1/2+\mathbb Z)/l_{AdS}$ (which includes the massless case in odd bulk dimension $d$) admit a real Euclidean pole in the propagator because of the presence of a logarithm. Therefore, the linear theory breaks conformal invariance and is unstable because of the presence of tachyons. This result is naturally understood if these Rarita-Schwinger fields belong to a supergravity multiplet. Half-integer masses then correspond to even-dimensional boundaries, where gravitational anomalies break conformal invariance \cite{Henningson:1998gx}. Also, the induced gravity theory obtained by imposing Neumann boundary conditions has been shown to be perturbatively unstable around the flat space boundary \cite{Compere:2008us}. The usual stability argument based on supersymmetry seems not to apply because of the presence of ghosts. On the contrary, masses $m = (d-2)/2l_{AdS}$ for even $d$ can be argued to lead to unitary and stable Neumann-type theories. Indeed, in this case, the Rarita-Schwinger fields belong to a supergravity multiplet and, as shown in \cite{Compere:2008us}, unitarity and stability hold for the linearized graviton around flat space.

One can ask how the results for $\cN = 1$ supergravity generalize to
theories with extended supersymmetry. We expect that the entire $\cN = 8$,
$d=4$ supergravity supermultiplet obtained from reducing $11d$
supergravity to $AdS_4 \times S^7$ will induce the $\cN = 8$
superconformal multiplet at the boundary. Indeed, the fact that $\cN = 8$
superconformal gravity can be coupled to the  Bagger-Lambert-Gustavsson
theory \cite{Gran:2008qx} seems to indicate that this is true. It has been
argued that supergravity on  $AdS_4 \times \mathbb C^4/\mathbb Z_k$ will
be only $\cN = 6$ supersymmetric for general $k$ in accordance with the
dual CFT \cite{Aharony:2008ug}. In that case, the $\cN = 6$ supergravity
multiplet is expected to induce the superconformal $\cN = 6$ multiplet at
the boundary.

It would be interesting to perform a similar analysis of boundary conditions for the vector and gravitino supermultiplets.  Another direction for future investigation is to follow the procedure of renormalizing the symplectic structure for spin-0, spin-1/2, and spin-1 fields. Scalar fields with $m^2 =-\frac{(d-1)^2}{4}+1$,  massive spin-1/2 fields with mass $m = 1/2$, and $U(1)$ gauge fields, might be particularly interesting because the Neumann theories do not appear to violate unitarity bounds.

Recently, the AdS/CFT correspondence has been applied as a mapping between strongly-coupled CFTs in the long wavelength (i.e., hydrodynamic) limit and AdS gravity, see e.g. \cite{Bhattacharyya:2008jc}.  One can further include  electromagnetic fields, leading to the magneto-hydrodynamics equations with fixed external sources \cite{Caldarelli:2008ze}.  Bulk spin-3/2 excitations have been interpreted in the CFT as ``phonino'' fluctuations \cite{Policastro:2008cx}. In these approaches, Dirichlet boundary conditions are enforced, implying that the boundary fields are non-dynamical.  It might be interesting, however, to explore this aspect of gauge/gravity duality with Neumann or mixed boundary conditions, where the boundary fields are also allowed to vary.

\begin{acknowledgments}
We are particularly grateful to Don Marolf for numerous valuable discussions and comments on an early draft of this manuscript. We also thank Stefan Hollands for discussions related to this work and Miranda Cheng, Marc Henneaux, Peter van Nieuwenhuizen and Kostas Skenderis for their valuable correspondence.  This work is supported in part by the US National Science Foundation under Grant No.~PHY05-55669, and by funds from the University of California.

\end{acknowledgments}

\appendix

\section{Conventions}
\label{sec:conventions}

Our conventions throughout this paper are as follows.  Greek letters $\mu, \nu, \lambda, \ldots$ denote spacetime indices and hatted letters $\hat \mu, \hat \nu, \hat \lambda \ldots = 0,1,2,\ldots$ denote indices on a flat internal space.  The flat-space gamma matrices satisfy $\{\gamma_{\hat \mu}, \gamma_{\hat \nu} \} = 2 \eta_{\hat \mu \hat \nu}$, where $\eta_{\hat \mu \hat \nu}$ is the metric of Minkowski space with signature $(-++\ldots)$.  The Hermitian conjugate of a gamma matrix is then $\gamma_{\hat \mu}^\dagger = \gamma_0 \gamma_{\hat \mu} \gamma_0$.  For a given spacetime metric $g_{\mu \nu}$, we can define an orthonormal frame $\{e^{\hat \mu}{}_{\mu} \}$ which satisfies $e^{\hat \mu}{}_\mu e^{\hat \nu}{}_\nu \eta_{\hat \mu \hat \nu} = g_{\mu \nu}$. We denote $e = \text{det}\; e^{\hat \mu}{}_{\mu} $.  The curved space gamma matrices are given by projecting with the orthonormal frame $\gamma_\mu = e^{\hat \nu}{}_\mu \gamma_{\hat \nu}$  and satisfy $\gamma_{(\mu} \gamma_{\nu )} = g_{\mu \nu}$.  We further define the antisymmetrized gamma matrices $\Gamma^{\mu_1 \ldots \mu_n} = \gamma^{[\mu_1} \ldots \gamma^{\mu_n]}$ and the Dirac conjugate of a spinor $\overline{\psi} = i \psi^\dagger \gamma^0$.  We assume that all spinors are anticommuting fields. Our conventions for the Riemann tensor are $R_{\mu\nu \hat \kappa \hat \rho} \equiv \d_\mu \omega_{\nu \hat \kappa \hat \rho} + \omega_{\mu \hat \kappa \hat \lambda}\omega_{\nu \;\; \hat \rho}^{\;\, \hat \lambda}-(\mu\leftrightarrow \nu)$ and $R_{\mu \hat \kappa} \equiv e^{\nu \hat \rho}R_{\mu\nu \hat \kappa \hat \rho }$, $R \equiv e^{\mu \hat \kappa} R_{\mu  \hat \kappa}$.

In four dimensions, Majorana fields obey the condition $\overline{\psi_\mu} = \psi^T_\mu C$, where $C$ is the charge conjugation matrix satisfying $C \gamma_\mu C^{-1} = -\gamma_\mu^T$ and $C^T = C^{-1} = C^\dagger = - C$.  We frequently use the flip identity $\overline{\psi_\nu} \gamma^{\mu_1} \ldots  \gamma^{\mu_n} \chi_\sigma = (-1)^n \, \overline{\chi_\sigma} \gamma^{\mu_n} \ldots  \gamma^{\mu_1} \psi_\nu$ for any Majorana fields $\psi_\nu, \chi_\sigma$. We define $\gamma_5 \equiv \gamma^1 \gamma^2 \gamma^3 i\gamma^0$, which anticommutes with all $\gamma_\mu$.  Our convention for the Levi-Civita tensor is $\varepsilon^{0123}=+1=-\varepsilon_{0123}$.  Our choice for the orientation of surfaces of constant radial coordinate is given by $\varepsilon^{\Omega ijk} = -\varepsilon^{ijk}$ in the coordinates defined by \eqref{metricd} or $\varepsilon^{rijk} = +\varepsilon^{ijk}$ in the coordinates defined by \eqref{asympt:metric}.   The tensor $\epsilon^{\hat \mu \hat \nu \hat \kappa \hat \rho}$ is given by $e^{-1} \varepsilon^{\hat \mu \hat \nu \hat \kappa \hat \rho}$.

\section{Asymptotic expansions}
\label{app:asymptsol}

The vielbein and the two projected Rarita Schwinger vector-spinors admit the expansions \eqref{assol} in the gauge \eqref{gauge:FG}. The inverse vielbein is $e^{i\,\hat i} = e^{i\,\hat i}_{(0)}e^{-r} -e^{i\,\hat i}_{(2)}e^{-3r}-e^{i\,\hat i}_{(3)}e^{-4r}+O(e^{-5r}),$ where $e^{i\,\hat i}_{(2)} \equiv e^{i}_{(0) \hat j} e_{(2)j}^{\hat j}e^{j \hat i}_{(0)}$ and similarly for $e_{(3)}$. Note that $e^{\hat i i}_{(2)} \equiv e_{(2)j}^{\hat i} g^{ij}_{(0)}$ is in general different than $e^{i\,\hat i}_{(2)}$. However, one can use the local Lorentz transformations \eqref{gauge3} to further fix the gauges $e_{(0)\hat i \, i}e^{\hat i}_{(2)j}=\frac 1 2 g_{(2)ij}$ and $e_{(0)\hat i \, i}e^{\hat i}_{(3)j}=\frac 1 2 g_{(3)ij}$, where $g_{(2)ij}$ and $g_{(3)ij}$ are the usual metric Fefferman-Graham coefficients
\begin{eqnarray}
\gamma_{ij} = e^{2r}g_{(0)ij} + g_{(2)ij} + e^{-r} g_{(3)ij} + O(e^{-4r})\, .
\end{eqnarray}

The extrinsic curvature is defined as $K_{ij} = \frac 1 2 \d_r \gamma_{ij}$. Expanding the definition of the spin connection \eqref{eq:spin}, we obtain the leading term
\begin{equation}
D^{(0)}_{[i} e_{(0)j]}^{\hat i} = \frac{\kappa^2}{4} \left( \bar \psi_{(0)i,+} \gamma^{\hat i} \psi_{(0)j,+}  \right) ,
\end{equation}
where the derivative $D^{(0)} = \d + \omega_{(0)}$ is associated with the leading part of the connection $\omega$. Expanding the spin connection in more detail, one can get the following expansions used in the main text,
\begin{eqnarray}
\omega^{\;\hat i \hat j}_r &=& e^{k[\hat i}\d_r e^{\hat j]}_k +\frac {\kappa^2}{2} \bar \psi^{[\hat i}_-\psi^{\hat j ]}_+= \frac{\kappa^2}{2}\bar \psi^{[\hat i}_{(2),-}\psi^{\hat j]}_{(0),+}e^{-2r}+\frac{\kappa^2}{2}\bar \psi^{[\hat i}_{(3),-}\psi^{\hat j]}_{(0),+}e^{-3r} +O(e^{-4r})\nonumber \\
\omega^{\;\hat j \hat r}_i &=& e^{k \hat j}K_{ik} - \frac{\kappa^2}{4}(\bar \psi_{i,-}\psi_+^{\hat j}-\bar \psi_{i,+}\psi_-^{\hat j}) \nonumber\\
&=& e^{\hat j}_{(0)i} e^r + e^{-r}(-e_{(2)i}^{\hat j} - \frac{\kappa^2}{4}(\bar \psi_{(2)i,-}\psi_{(0)}^{\hat j}-\bar \psi_{(0)i,+}\psi_{(2),-}^{\hat j})     )\label{spinconnection}\\
&&+ e^{-2r}(-2e_{(3)i}^{\hat j} - \frac{\kappa^2}{4}(\bar \psi_{(3)i,-}\psi_{(0)}^{\hat j}-\bar \psi_{(0)i,+}\psi_{(3),-}^{\hat j})     )+O(e^{-3r}) \nonumber\\
\omega^{\;\hat i \hat j}_i &=& e^{0r}\omega^{\hat i \hat j}_{(0)i}+O(e^{-2r})\nonumber
\end{eqnarray}
The expansions of the Ricci tensor and the stress-tensor can be obtained straightforwardly, but are not given here.

Lastly, we record several results that were used in the main text. For the Rarita-Schwinger field, the form of the Fefferman-Graham coefficient $\psi_{(4)i,+}$ is given by
\begin{eqnarray}
\psi_{(4)i,+} = \frac 1 3 \slash \hspace{-0.7em} D_{(0)} \psi_{(3)i,-}+ \frac{1}{2}e^{\hat i}_{(3)i} \gamma^j \gamma_{\hat i}\psi_{(0)j,+} + (\psi_{(0),+}\cdot \psi_{(0),+}\cdot \psi_{(3),-})\text{-terms}\,,\label{psi4}
\end{eqnarray}
where $(\psi_{(0)}\cdot \psi_{(0)}\cdot \psi_{(3)})\text{-terms}$ are higher order terms linear in $\psi_{(3),-}$ and its first derivative and bilinear in $\psi_{(0)i,+}$ and its first derivative.
For the supersymmetry-generating parameter $\epsilon$, we note
\begin{eqnarray}
\eps_{(3),+}[\eps_-,\eps_+,\xi^r] &=& -\frac{\kappa^2}{4}(\bar \eps_- \psi_{(0),+}^i - \bar \eps_+ \psi_{(2),-}^i) \psi_{(0)i,+} - \frac{\kappa}{2}\d^i \xi^{\hat r} \psi_{(0)i,+}+\frac{\kappa^2}{16}\psi^{[\hat i}_{(2),-}\psi^{\hat j]}_{(0),+}\Gamma_{\hat i\hat j}\eps_+,\nonumber \\
\eps_{(4),-}[\eps_-,\eps_+,\xi^r] &=& -\frac{\kappa^2}{4}(\bar \eps_- \psi_{(0),+}^i - \bar \eps_+ \psi_{(2),-}^i) \psi_{(2)i,-} -\frac{\kappa}{2}\d^i \xi^{\hat r} \psi_{(2)i,-} + \frac{\kappa^2}{16}\psi^{[\hat i}_{(2),-}\psi^{\hat j]}_{(0),+}\Gamma_{\hat i\hat j}\eps_-  \nonumber \\
\eps_{(4),+}[\eps_+] &=& \frac{\kappa^2}{6}\bar \eps_+ \psi_{(3),-}\psi_{(0)i,+}+\frac{\kappa^2}{24}\bar\psi_{(3),-}^{[\hat i} \psi_{(0),+}^{\hat j]}\Gamma_{\hat i\hat j}\eps_+\,.\label{eps4}
\end{eqnarray}

\providecommand{\href}[2]{#2}\begingroup\raggedright\endgroup

\end{document}